\documentclass[aps,prx,reprint,superscriptaddress,nofootinbib,floatfix]{revtex4-2}

\usepackage[utf8]{inputenc}
\usepackage[T1]{fontenc}
\usepackage{amsmath,amssymb,amsthm,amsfonts,bm}
\usepackage{physics}
\usepackage{graphicx}
\usepackage{xcolor}
\usepackage{hyperref}
\hypersetup{
 hidelinks,
 pdftitle={Trial-space co-design for quantum simulation on noisy hardware},
 pdfauthor={Bo Peng, Vibin Abraham, and Himadri Pathak},
 pdfsubject={Research manuscript on noise-aware quantum simulation},
 pdfkeywords={noise-aware co-design, quantum simulation, quantum hardware, quantum subspace diagonalization}
}
\usepackage{enumitem}
\usepackage{booktabs}

\newcommand{\Span}{\operatorname{span}}

\begin{document}

\title{Trial-space co-design for quantum simulation on noisy hardware}

\author{Bo Peng}
\email{peng398@pnnl.gov}
\author{Vibin Abraham}
\affiliation{Integrated Discovery Sciences Directorate, Pacific Northwest National Laboratory, Richland, Washington 99354, USA}
\author{Himadri Pathak}
\affiliation{Quantum Mathematical Science Team, Division of Applied Mathematical Science, RIKEN Center for Interdisciplinary Theoretical and Mathematical Sciences (iTHEMS), 2-1 Hirosawa Wako, Saitama 351-0198, Japan}
\affiliation{Computational Molecular Science Research Team, RIKEN Center for Computational Science (R-CCS), 7-1-26 Minatojima-minami-machi, Chuo-ku, Kobe, Hyogo 650-0047, Japan}

\begin{abstract}
Quantum simulations on noisy hardware are often designed as if the machine were ideal, with device limitations considered only after the basis and measurement plan have been chosen. We instead use information about the physical problem and the device at an earlier stage, to decide which states and evolution times should form a compact simulation basis. Because the target problem remains unchanged, the method adapts only the representation used to solve it. Tests on many-body and molecular models show that this approach can reveal useful directions that ordinary time evolution misses, while also exposing an important limitation: some of these directions can become too similar to estimate reliably. Realistic device data further show that hardware response can change the preferred states and evolution times before execution. Across seven acquisition days, a four-qubit experiment shows a median 54\% reduction in projected-energy error from problem-informed time choices relative to uniform sampling under matched hardware conditions. These results provide a proof of principle for simulation-basis design as a practical interface between quantum algorithms and the hardware that executes them.
\end{abstract}

\maketitle

\section{Introduction}

Quantum algorithms are usually designed around an ideal description of the machine. The physical device enters later, when calibration information is used to choose a qubit mapping, compile circuits, design measurements, or mitigate errors~\cite{Alexeev2021CoDesign,Wittler2021C3,JavadiAbhari2024Qiskit}. This separation is natural, but it leaves an important question open: \emph{should what we know about the real device also influence how the simulation itself is represented?}

The question is particularly important when a large quantum problem is reduced to a much smaller set of states. In many-body simulation, quantum chemistry, lattice gauge theory, and related problems, the full state space quickly becomes too large to work with directly~\cite{Miessen2023QuantumDynamics,Fauseweh2024ManyBody,Battaglia2024ActiveSpace,Pardo2023Gauge,Weimer2021OpenSystems}. A calculation therefore often uses a compact \emph{trial space}: a set of states chosen to retain the physics needed for the target quantity. The quality of the calculation depends not only on how accurately those states can be measured, but also on whether the right physical directions were included in the first place. Information missing from the chosen space cannot be recovered later by improving the measurement or correcting its noise.

This representation problem is broader than any particular quantum algorithm. Variational ans\"atze, configuration- or operator-generated bases, reduced models, and time-evolved states all make choices about which part of a much larger problem will actually be represented on the device. Projected quantum simulations provide a useful setting in which to study this choice: selected states are prepared, their mutual overlaps and Hamiltonian matrix elements are measured, and the physical problem is solved within the resulting small space~\cite{McClean2017QSE,Epperly2022QSD,Shen2023RTK}.

Many strategies already improve how such spaces are constructed or used. Candidate states can be generated by real-time evolution, filtering, imaginary-time or Lanczos procedures, moment expansions, and sampled or multireference constructions~\cite{Parrish2019QFD,Cohn2021QFD,Motta2020QITE,YeterAydeniz2020QLanczos,Jones2022QCM,Klymko2022RealTime,Cortes2021QKSD,Yoshioka2025Krylov,Yu2025SKQD,Stair2020MRSQK,Kirby2023Lanczos}. Adaptive selection, pruning, measurement design, and regularization can then make these spaces smaller, more accurate, or more stable~\cite{Prudhomme2002ReducedBasis,Rammal2026Orthogonal,Asthana2025QKUD,Getelina2024HardwareNoise,OLeary2025PQSE,DeVriendt2026Angle,Lee2024Sampling,Kirby2024Errors,Zhang2024MeasEff,Lee2025Sampling,Oliveira2026Instabilities}. Related work has also used hardware noise to select useful subspace operators~\cite{Getelina2024HardwareNoise}, engineered wavefunction structure to improve sampling~\cite{Xu2026FilterAssisted}, demonstrated Krylov calculations on large processors~\cite{Yoshioka2025Krylov}, and used device calibration to improve compilation and control~\cite{Wittler2021C3,JavadiAbhari2024Qiskit}.

Here we ask a different question: can information about a physical response or a real device help choose the compact representation \emph{before} the measurements needed for the reduced calculation are made? We call this \emph{trial-space co-design}. The target Hamiltonian and the physical quantity being computed are left unchanged. What changes is the set of states, evolution times, or hardware mappings used to represent that problem.

Two forms of information enter this design. First, a reproducible state-dependent response can reshape candidate states. Unlike closed time evolution, which changes relative phases while preserving spectral weights, such a response can also change relative amplitudes and thereby expose directions that would otherwise require several time-evolved states to reconstruct. Second, information from a real processor can identify which mappings and evolution times remain practical under its measured coherence and error profile. In both cases, information that would normally enter only after the algorithm is specified is used earlier to decide what should be represented and measured.

\begin{figure*}[t]
\centering
\includegraphics[width=0.97\textwidth]{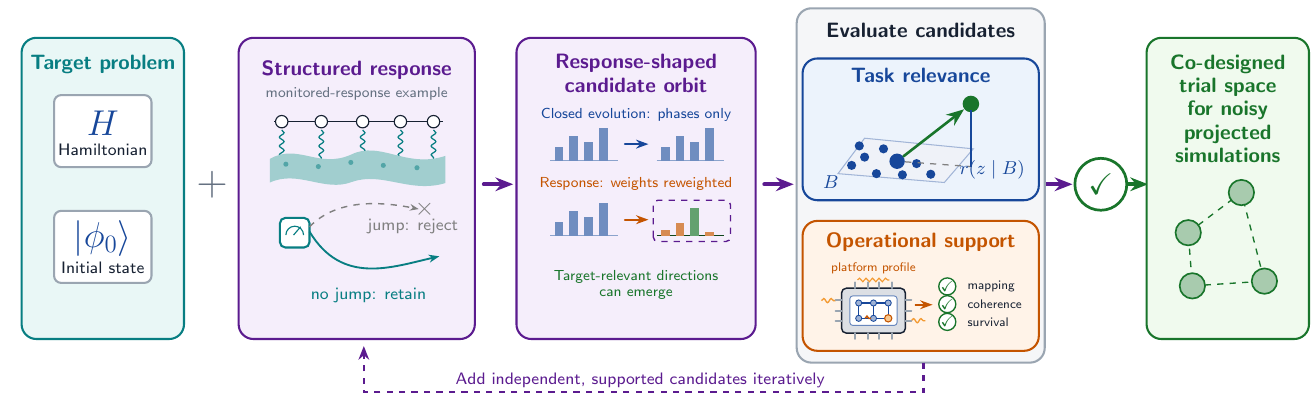}
\caption{\textbf{Trial-space co-design uses structured response and platform information for noisy projected quantum simulations.} From left to right, the target Hamiltonian, initial state, and an accessible structured response define a response-shaped candidate orbit. Closed evolution changes phases while preserving component weights, whereas the response can also reweight those components and expose target-relevant directions. The jump/no-jump record illustrates one monitored-response implementation of this candidate-generation step. Candidates are then screened separately for task relevance and operational support. Here \(B\) is the current trial space and \(r(z\mid B)\) measures the missing target information addressed by candidate \(z\). Operational support uses a noisy-platform profile as design information to check mapping, coherence, and survival. The platform profile therefore acts as a design screen rather than as a second state measurement. Candidates that are independent, relevant, and supported are added iteratively to the co-designed trial space for noisy projected simulations. The resulting compact space directs projected-matrix acquisition toward coordinates that carry missing target information while remaining distinguishable and implementable under device noise. The target Hamiltonian and physical objective remain unchanged.}
\label{fig:codesign_overview}
\end{figure*}

Figure~\ref{fig:codesign_overview} summarizes the resulting workflow. The target problem and a structured response first define a response-shaped candidate orbit. Because the candidate generator is a monitored response, we use a jump/no-jump record to show how conditional evolution can reshape the state. The candidates are then evaluated for task relevance and operational support. In particular, mapping, coherence, and survival determine whether a coordinate remains viable on the device, so a measured platform profile, as design information, screens these quantities before acquisition. Independent candidates that pass both evaluations are added iteratively, yielding a co-designed trial space whose coordinates remain distinguishable and implementable on noisy hardware. Thus, the workflow separates the response that reshapes candidate states from the platform information that determines whether those states can support the subsequent noisy simulation. In both roles, the useful input is reproducible structure rather than uncontrolled noise or drift.

This construction is related to, but distinct from, using nonunitary dynamics to prepare a target state. Imaginary-time, Lanczos, and filtering methods already use amplitude reweighting, and filter-assisted subspace methods can engineer a wavefunction to improve sampling~\cite{Xu2026FilterAssisted}. Dissipative preparation uses an open system to drive the system toward the desired state~\cite{Verstraete2009Dissipation,Li2025Dissipative}, while Liouvillian Krylov methods study the dynamics or spectrum of an open system~\cite{HerreraMarti2026OpenKrylov}. Here, structured response generates a family of trial states whose usefulness is assessed through matched selection, overlap conditioning, and preparation support, while the original Hamiltonian remains the target of the calculation.

We test this idea from mechanism to hardware. A Hubbard model first shows how a monitored local response can add useful trial-space directions, while also revealing an important limitation: directions that are physically informative can become too similar to estimate reliably. Transverse-field Ising simulations then test whether the same mechanism transfers when a different local physical response is relevant. Molecular N\(_2\) calculations provide a complementary test: there the response is constructed from the molecular Hamiltonian rather than from an assumed environment, showing that the design workflow can extend to a chemically distinct representation without requiring a universal response model or a universal performance gain. Archived processor data next show that measured device properties can change the preferred mapping and evolution times. Finally, repeated experiments on real hardware test whether choosing evolution times using the target problem improves the measured energy relative to uniform sampling under otherwise matched conditions.


\section{Results}

\subsection{How structured response changes a simulation representation}

Our first question is how a reproducible response can change a trial space at all. The answer comes from comparing what two candidate-state generators can change. Closed evolution rotates quantum phases but preserves the weights of the energy components. A structured response can also reweight those components. When that reweighting favors target-relevant content, it can supply a useful direction that closed evolution must synthesize from several time samples. We first make this contrast precise, then test whether the added direction improves accuracy without destroying numerical stability.

Let \(H\) denote the target Hamiltonian and \(\{\ket{E_b}\}\) its energy eigenstates. Any chosen initial state can be written as
\(
 \ket{\Phi_0}=\sum_b c_b\ket{E_b}.
\)
After a real-time interval \(t\), each energy component acquires a phase,
\(
 e^{-iHt}\ket{\Phi_0} = \sum_b c_b e^{-iE_bt}\ket{E_b}.
\)
The equations make the limitation precise: the relative phases change, but the populations \(\lvert c_b\rvert^2\) do not. Every state along the closed-time orbit therefore has the same spectral weight distribution. A compact basis built only from closed-time samples must recover the target low-energy direction by combining several phase-rotated copies of that same distribution.

To move beyond phase rotation, we introduce a structured open-system response that can change relative amplitudes in a reproducible way. Let \(L_\ell\) denote monitored local processes and \(R=\sum_\ell L_\ell^\dagger L_\ell\) their positive rate operator. The key design choice is how \(R\) is specified from accessible model structure. It need not be diagonal in the energy basis and need not require the eigensystem of \(H\). In other words, the amplitude reweighting becomes a controlled state-generation step when the response is monitored. Suppose that no transition is detected during a finite time interval, an outcome we call a \emph{no-jump record}. Components on which \(R\) is large are then suppressed more strongly. When the longer-lived directions contain more of the target low-energy content, the record can generate a useful trial direction directly rather than reconstructing it from many phase-rotated closed-time states. This leads to the design principle: \textit{structured, state-dependent attenuation can reshape the trial space toward target-relevant directions during construction, before complete matrix acquisition and the final reduced solve.} The useful ingredient is therefore the reproducible information carried by the response, not noise of arbitrary form or strength. This is the first part of the co-design principle: \textit{a characterized response can enlarge the set of directions offered to the algorithm before the final measurement problem is fixed.}

The short-time expression makes this geometric statement precise. With \(\langle R\rangle_0=\langle\Phi_0|R|\Phi_0\rangle\), normalization gives
\begin{equation}
 \ket{\psi_{\mu,t}}
 =\ket{\Phi_0}-itH\ket{\Phi_0}
 -\frac{\mu t}{2}(R-\langle R\rangle_0)\ket{\Phi_0}
 +O(t^2).
 \label{eq:response_tangent}
\end{equation}
The first two terms lie in the local space already generated by the initial state and ordinary Hamiltonian evolution. The response term can add a third direction. More precisely, let \(P_0\) project onto \(\Span\{\ket{\Phi_0},H\ket{\Phi_0}\}\). If \((I-P_0)R\ket{\Phi_0}\neq0\), part of the response lies outside that local closed-time space and supplies a genuinely new first-order direction. This geometric condition does not by itself guarantee a lower projected energy or stable estimation, but the selection and the overlap spectrum determine whether the added direction is usable. The full statement and proof are given in Supplementary Section~1.2.

For a finite time interval, we label each candidate by a response setting \(z=(\mu,t)\), where \(\mu\) controls the attenuation strength and \(t\) controls its duration. The operator for obtaining no detected transition during this interval is
\begin{equation}
 K_z \equiv K_{\mu,t}
 =\exp\!\left[t\left(-iH-\frac{\mu}{2}R\right)\right].
 \label{eq:no_jump_operator}
\end{equation}
In quantum-trajectory language, \(K_z\) is the Kraus operator for the no-jump record~\cite{Dalibard1992WaveFunction}. The target Hamiltonian supplies the coherent evolution, while \(R\) supplies state-dependent attenuation. Only in the spectral-reference case, where \(R=\sum_b\gamma_b\ket{E_b}\!\bra{E_b}\) commutes with \(H\), does Eq.~\eqref{eq:no_jump_operator} reduce to \(\sum_b e^{-iE_bt-\mu\gamma_bt/2}\ket{E_b}\!\bra{E_b}\). The main candidate generators below instead construct \(R\) from local operators or sparse Hamiltonian data. Varying \((\mu,t)\) then changes both phases and amplitudes while starting from one common input state.

Applying \(K_z\) and normalizing the retained branch produces the candidate associated with that no-jump record,
\begin{equation}
 \ket{\psi_z}
 =
 \frac{K_z\ket{\Phi_0}}{\sqrt{p_z}},
 \qquad
 p_z
 =
 \bra{\Phi_0}K_z^\dagger K_z\ket{\Phi_0}.
 \label{eq:conditional_branch}
\end{equation}
The quantity \(p_z\) is the squared norm of the unnormalized branch and the probability of obtaining the corresponding no-jump record. Dividing by \(\sqrt{p_z}\) normalizes the candidate. This probability creates an operational trade-off: stronger attenuation may produce a more distinctive candidate, but it may also make that candidate less likely to prepare. Under independent repetitions, the expected number of attempts needed to obtain it scales as \(1/p_z\). See Supplementary Sections~1.1--1.2 for details.

The construction is not tied to one physical realization. Continuous monitoring can implement the no-jump map directly, while an ancilla and postselection can produce the same map. The Hubbard example below uses the latter construction. Averaging over all outcomes would instead give the usual unconditional open-system state and erase the record-specific reweighting that creates the candidate.

Candidate generation is only the first stage. The response settings produce a library of possible directions but do not determine which ones should enter the trial space. Some candidates add useful information, whereas others are nearly redundant or too unlikely to prepare. We therefore apply one fixed selector to balance low-energy relevance, linear independence, and preparation support, as summarized in Figure~\ref{fig:codesign_overview}. Starting from the common \(t=0\) state, this iterative procedure adds the highest-scoring candidate and then updates the trial space before scoring the remaining candidates again.

Task relevance is evaluated relative to the current trial space. At each step, that space gives an approximate low-energy state. Acting with \(H\) exposes the part of the target equation that the current space cannot yet represent, which defines the residual. The selector favors candidates aligned with this residual, penalizes near dependence, and weights each candidate by an implemented branch-derived quality factor. In the explicit matched-selector controls, the same selection rule is applied to both candidate libraries, so differences in the selected trial space are not created by retuning the selector. For small-system simulations, one can use exact candidate vectors, \(H\)-actions, residuals, Gram information, and projected energies as controlled proxies for this score. They test the design principle in classically tractable systems rather than supply a final scalable selector. For larger implementations, these quantities must be estimated or replaced with measurable surrogates, reduced response models, or adaptive measurements; the associated measurement and optimization cost is not characterized here. Supplementary Section~1.3 gives the corresponding information requirements.
After selection, a basis
\(
 B=\{\ket{\psi_{z_1}},\ldots,\ket{\psi_{z_m}}\}
\)
defines the ordinary projected Hamiltonian problem,
\(
 (H_B)_{ij}
 =
 \bra{\psi_{z_i}}H\ket{\psi_{z_j}}.
\)
The response thus determines the working basis, not the physical objective. The Hamiltonian, symmetry sector, and target energy remain unchanged, and only the coordinates used to approximate them are adapted.

With candidate generation, selection, and projection now separated, the remaining results move from mechanism to transfer and then to operational realization. We begin with a Hubbard benchmark that separates the original end-to-end comparison from a matched-selector control at fixed target Hamiltonian, input state, and accepted dimension.


\subsection{Structured local response improves a compact Hubbard simulation}

\begin{figure*}[t]
\centering
\includegraphics[width=0.95\textwidth]{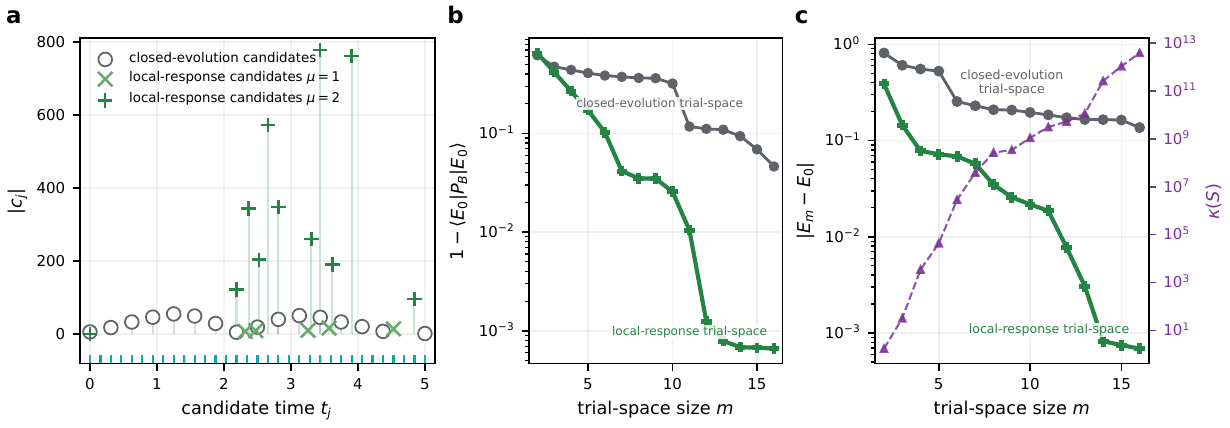}
\caption{\textbf{Local charge response reshapes the Hubbard trial space.} The six-site, half-filled model has \(U/t=16\). Candidates use Eq.~\eqref{eq:no_jump_operator} with \(R_{\rm loc}\) from Eq.~\eqref{eq:local_hubbard_response} and require no target eigensystem. \textbf{a,} Coefficients of the lowest-energy projected states at matched trial-space size \(m=16\). Horizontal position gives candidate time, vertical position gives \(\lvert c_j\rvert\), and teal ticks mark the available time grid. Crosses and plus signs distinguish response strengths \(\mu=1\) and 2. The common \(t=0\) state is independent of \(\mu\). The projected energies are \(-0.9239\) for closed evolution and \(-1.0597\) for local response, compared with exact energy \(-1.0604\). The nonorthogonal coordinates satisfy \(c^\dagger S c=1\), so \(\lvert c_j\rvert^2\) are not probabilities. \textbf{b,} Fraction of exact ground-state weight outside the selected trial space. At \(m=16\), the closed and response-informed values are \(4.63\times10^{-2}\) and \(6.61\times10^{-4}\). \textbf{c,} Projected-energy error and overlap condition number. The corresponding errors are \(1.37\times10^{-1}\) and \(6.88\times10^{-4}\). Supplementary Table~S1 applies the same selector to both candidate libraries: the response span is 32.9-fold better before truncation, but at the fixed \(10^{-3}\) overlap floor it retains only three modes and has error 0.1312, compared with nine modes and error 0.08835 for selected closed evolution.}
\label{fig:hubbard_representation}
\end{figure*}

We use a six-site periodic Hubbard model at half filling and \(U/t=16\). Six sites give a 400-dimensional fixed-particle sector that retains nontrivial many-body charge fluctuations while still allowing exact diagnostics for the matched controls. The ratio \(U/t=16\) places the model in the strongly interacting regime, where doublon--hole configurations are energetically costly. Local charge fluctuation is therefore a physically natural response coordinate and provides a stringent test of whether amplitude shaping can improve a compact trial space. The main candidate generator monitors
\begin{equation}
 L_i=n_{i\uparrow}+n_{i\downarrow}-1,
 \qquad R_{\rm loc}=\sum_iL_i^\dagger L_i .
 \label{eq:local_hubbard_response}
\end{equation}
Thus \(R_{\rm loc}\) vanishes on singly occupied configurations and assigns weight two to every doublon--hole pair. The operator \(R_{\rm loc}\) depends only on local occupations and can therefore be constructed without diagonalizing the target Hamiltonian. We sample evolution time and response strength over a range set by the local operator scale and time resolution.

Figure~\ref{fig:hubbard_representation} first compares the complete conventional and response-informed workflows. The prescribed closed-time construction samples a phase-only orbit, whereas the adaptive local-response construction selects from candidates whose amplitudes can also change. Both start from the same input state at \(t=0\). Figure~\ref{fig:hubbard_representation}a shows the lowest-energy state after projecting \(H\) into each trial space at \(m=16\), \(\lvert\Psi_m\rangle=\sum_jc_j\lvert\psi(t_j,\mu_j)\rangle\). The response changes both the selected evolution times and the coefficients used to combine their candidates. The projected energies are \(-0.9239\) for the closed-evolution trial space and \(-1.0597\) for the local-response trial space, compared with the exact value \(-1.0604\). Note that, because the candidate states are not orthogonal, the individual \(\lvert c_j\rvert^2\) should not be read as probabilities.

Figure~\ref{fig:hubbard_representation}b asks how much of the exact ground state lies outside each trial space. We quantify this missing fraction as
\(
\epsilon_{\rm span}=1-\langle E_0|P_B|E_0\rangle,
\)
where \(P_B\) is the orthogonal projector onto the selected span. The exact ground state is used only for evaluation, not for candidate generation or selection. At \(m=16\), the missing fraction decreases from \(4.63\times10^{-2}\) for closed evolution to \(6.61\times10^{-4}\) for the local-response construction, a roughly 70-fold reduction. The corresponding projected-energy error decreases from \(1.37\times10^{-1}\) to \(6.88\times10^{-4}\), about a 200-fold reduction (Figure~\ref{fig:hubbard_representation}c). These values compare the complete adaptive response workflow with prescribed closed-time sampling.

Applying the same selector to the closed-time candidate library moves both diagnostics partway toward the response-informed result: the energy error becomes \(2.265\times10^{-2}\) and the missing target weight \(2.859\times10^{-2}\) (Supplementary Section~2.3 and Table~S1). Thus, response-shaped candidate generation retains an approximately 33-fold energy-error advantage under the same selector before overlap truncation, showing that adaptive selection alone does not account for the gain. Supplementary Fig.~S1 separates amplitude reweighting from time-coordinate selection.

That gain, however, is limited by numerical distinguishability of the selected coordinates. The response-shaped basis is much more ill-conditioned than the matched selected closed-time basis and retains only three stable overlap modes at the fixed \(10^{-3}\) floor, compared with nine for closed evolution. After this truncation, the response advantage is no longer retained; the same trend appears across the selector-sensitivity study and in the held-out \(U/t=12\) control (Supplementary Section~2.3 and Table~S1). Thus the response can expose highly informative low-energy directions that are nevertheless too nearly dependent to use directly in a stable projected solve. Representation compression and estimable compression are therefore distinct requirements.

For reference, an eigensystem-derived response lowers the endpoint error only further to \(3.24\times10^{-4}\), showing that the local occupation response captures much of the available representational gain without requiring the target eigensystem (Supplementary Section~2.1).


\subsection{Response-informed trial-space design across physical problems}

\begin{figure*}[t]
 \centering
 \includegraphics[width=0.97\textwidth]{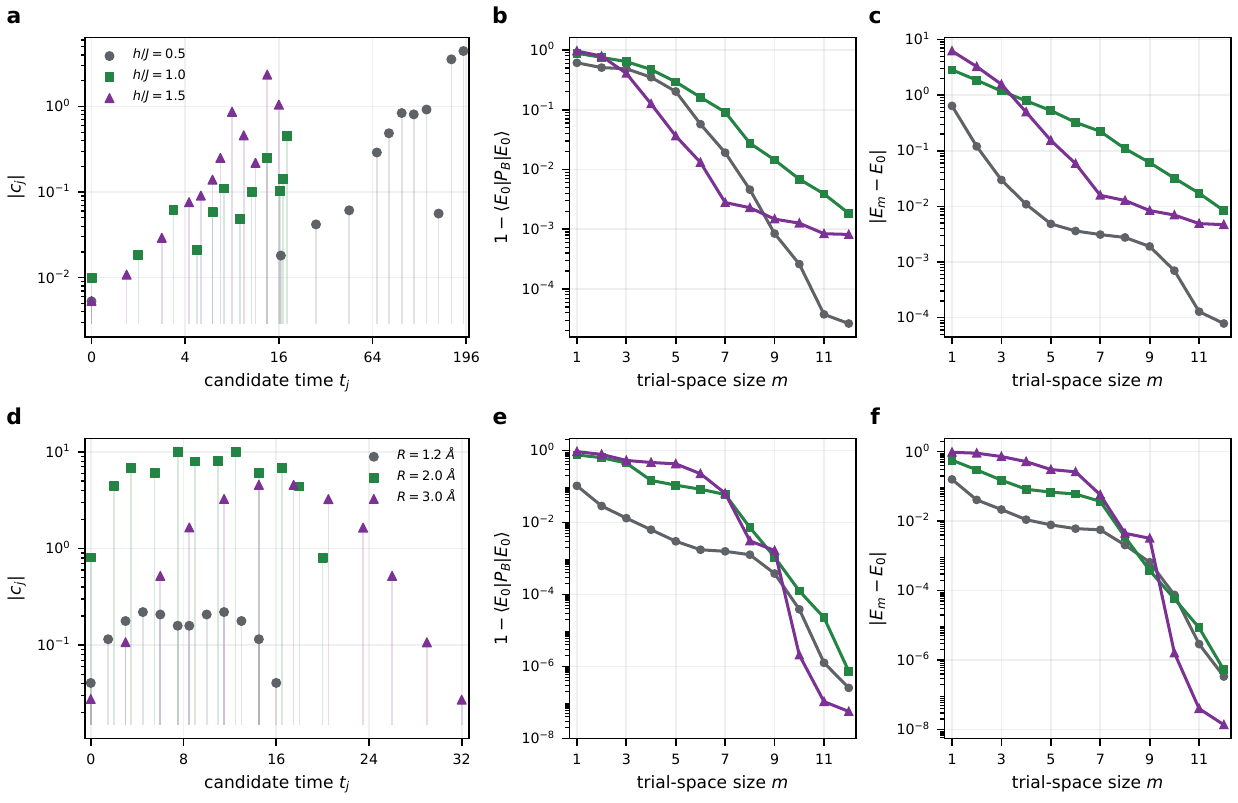}
 \caption{\textbf{System-dependent response designs extend the trial-space
 construction across physical problems.} The top row shows the TFIM at
 \(h/J=0.5,1.0,1.5\), and the bottom row shows N$_2$ at
 \(R=1.2,2.0,3.0\)~\AA. \textbf{a,d,} Magnitudes of the Ritz coefficients
 in the selected coordinates, with \(c^\dagger S c=1\).
 \textbf{b,e,} Target weight missing from successive prefixes of the selected
 trial space. \textbf{c,f,} Corresponding projected-energy error in units of
 \(J\) for TFIM and Hartree for N$_2$. The target is the lowest
 input-supported eigenstate; for stretched N$_2$ at \(R=3.0\)~\AA,
 this is eigenstate index three rather than the absolute ground state.
 All calculations use \(m=12\). TFIM candidates use
 Eq.~\eqref{eq:tfim_local_response}, whereas N$_2$ uses the
 Hamiltonian-derived response in Eq.~\eqref{eq:molecular_response_score}.
 Matched closed-evolution controls show clear gains for TFIM but no uniform
 advantage for N$_2$; the molecular example therefore tests transfer of the
 design strategy rather than universal improvement. Candidate generation and
 selection use no target eigensystem; exact target information enters only
 the reported evaluation metrics.}
 \label{fig:cross_model_representation}
\end{figure*}

The Hubbard result shows that a local response can expose useful trial-space directions, but the appropriate response need not be the same for another physical problem. Different Hamiltonians emphasize different local penalties, configurations, and timescales. We therefore ask whether the same representation-design strategy remains useful when the response is constructed from the structure available in each system.

We consider two complementary examples. The transverse-field Ising model (TFIM) retains a local lattice structure, but the relevant excitations are spin defects rather than charge fluctuations. It therefore tests whether a physically motivated local response can be adapted to a different many-body Hamiltonian. N$_2$ provides a more distinct case: its trial states are electronic configurations, and no physical monitoring process is assumed. Instead, we construct an algorithmic response directly from the molecular Hamiltonian. The molecular example therefore tests the transfer of the design strategy rather than of a particular physical response.

For the open-boundary TFIM, we use the dominant local energetic penalty to define
\begin{equation}
 R_{\rm TFIM}=\begin{cases}
 \displaystyle\sum_{i=1}^{N_s-1}\frac{I-Z_iZ_{i+1}}{2},&h/J<1,\\[4pt]
 \displaystyle\sum_{i=1}^{N_s}\frac{I-X_i}{2},&h/J\geq1.
 \end{cases}
 \label{eq:tfim_local_response}
\end{equation}
For \(h/J<1\), this operator counts domain walls, which are costly in the interaction-dominated regime. For \(h/J\geq1\), it counts spins misaligned with the transverse field. In either case, the response suppresses locally unfavorable configurations more strongly when used in Eq.~\eqref{eq:no_jump_operator}. The construction depends only on local Hamiltonian structure and does not require target eigenvalues or eigenvectors.

The molecular problem requires a different definition. For a configuration \(\lvert D_j\rangle\), the diagonal matrix element \(\epsilon_j=H_{jj}\) provides a local energy estimate, while a nonzero off-diagonal element \(H_{jk}\) identifies a directly connected rearrangement. We combine this information into
\begin{equation}
 \gamma_j=\sum_{k\in\mathcal C(j),\,\epsilon_k<\epsilon_j}
 |H_{kj}|^2W(\epsilon_j-\epsilon_k),
 \label{eq:molecular_response_score}
\end{equation}
and define \(R_{\rm mol}=\sum_j\gamma_j\lvert D_j\rangle\!\langle D_j\rvert\). Configurations with strong couplings to lower-energy alternatives are therefore attenuated more rapidly. Unlike the TFIM construction, this response is inferred from the Hamiltonian rather than from a monitored environment and should be viewed as an algorithmic filtering surrogate, not as a physical decay model.

For each system, the response strength and time window are selected together subject to two feasibility requirements: the retained branch must remain sufficiently supported, and the resulting coordinates must remain well-conditioned. Among feasible designs, we retain the one with the lowest projected Ritz energy. This projected quantity is the task objective itself; the exact target eigenstate is used only after selection to evaluate the result. The full candidate grids and sensitivity tests are given in Supplementary Section~4 and Figs.~S3--S4.

Figure~\ref{fig:cross_model_representation} shows how the selected coordinates
build the trial spaces. For TFIM, the preferred response strength and time
window change substantially across the three field regimes, reflecting the
different local excitations that control the low-energy physics. Despite
these different choices, the selected \(m=12\) spaces reach fidelities between
0.9972 and 0.99986 and small projected-energy errors across all three cases.
Matched closed-evolution controls show that the response-informed construction
provides the clearest improvement in this lattice example.

N$_2$ provides a complementary transfer test. The selected time window grows from \(T=16\) near equilibrium to \(T=32\) at the stretched geometry, while the response strength remains weak. Across all three geometries, the selected spaces recover their input-supported targets with fidelities above \(0.999999\) and energy errors well below chemical accuracy. These results show that response-informed trial-space design extends to a chemically distinct representation when the response is constructed from available Hamiltonian structure rather than an assumed physical environment.

These examples establish the part of the framework that depends on the physical problem itself. The response form, strength, and useful time coordinates all change with the Hamiltonian, but the design logic is the same: use accessible structure to propose candidate directions, retain those that are both useful and numerically viable, and solve the original Hamiltonian in the resulting space. The next step is to introduce information that comes not from the model, but from the processor on which the calculation will be run.


\subsection{Problem and platform information jointly shape trial-space design}

The preceding examples construct the response from the physical problem itself. We now add a second source of information: the properties of the processor on which the calculation will be run. The target Hamiltonian still determines which trial space is useful, while the platform profile constrains which mappings and evolution times are operationally supported. Trial-space design can therefore combine what the problem requires with what the device can reliably implement.

We test this using the same four-site TFIM while introducing an archived processor profile containing connectivity, gate timing and errors, coherence, and readout information. Each candidate evolution time and logical-to-physical mapping then defines both a trial state and a concrete native schedule. We retain six coordinates that satisfy routing, coherence, response-support, and conditioning requirements, and select among them using the projected Ritz energy. The target eigenstate is not used in this design step; implementation details and the complete response model are given in Supplementary Section~5.

A matched response-free sweep over the same candidate dictionary isolates the effect of adding platform response. The response-free design selects the uniform times \(\{0,0.5,1.0,1.5,2.0,2.5\}\), whereas the platform-aware design selects \(\{0,0.5,0.75,1.25,1.5,2.0\}\) together with a different mapping. At \(m=6\), the latter lowers the exact span error from \(3.64\times10^{-4}\) to \(9.75\times10^{-5}\) and the projected-energy error from \(7.21\times10^{-4}J\) to \(2.72\times10^{-4}J\). Platform information therefore changes the working representation before simulated acquisition noise is introduced.

A circuit-level simulation then asks whether this redesigned representation remains useful when the projected matrices must be estimated from finite-shot, noisy circuits. Using shallow local measurements, the complete platform-aware workflow reaches an \(m=6\) energy error of \(6.93\times10^{-4}J\), compared with \(5.37\times10^{-3}J\) for the conventional noisy workflow. This is an end-to-end comparison because the two acquisition pipelines also differ in their calibration-derived treatment; the matched exact comparison above provides the cleaner evidence for the representation-design effect. Supplementary Section~5.1, Fig.~S5, and Tables~S2--S3 report the complete archived-device controls.

Additional archived-device tests apply the same design procedure on larger coupling graphs; Supplementary Section~5.2 and Table~S4 report four-site transfer, while Supplementary Section~5.3 and Fig.~S6 report the eight-site design and schedule-support projection. We next replace archived platform models with a prospective experiment on a live processor.


\subsection{Problem-informed time coordinates improve projected-energy accuracy on real hardware}

\begin{figure*}[t]
\centering
\includegraphics[width=0.98\textwidth]{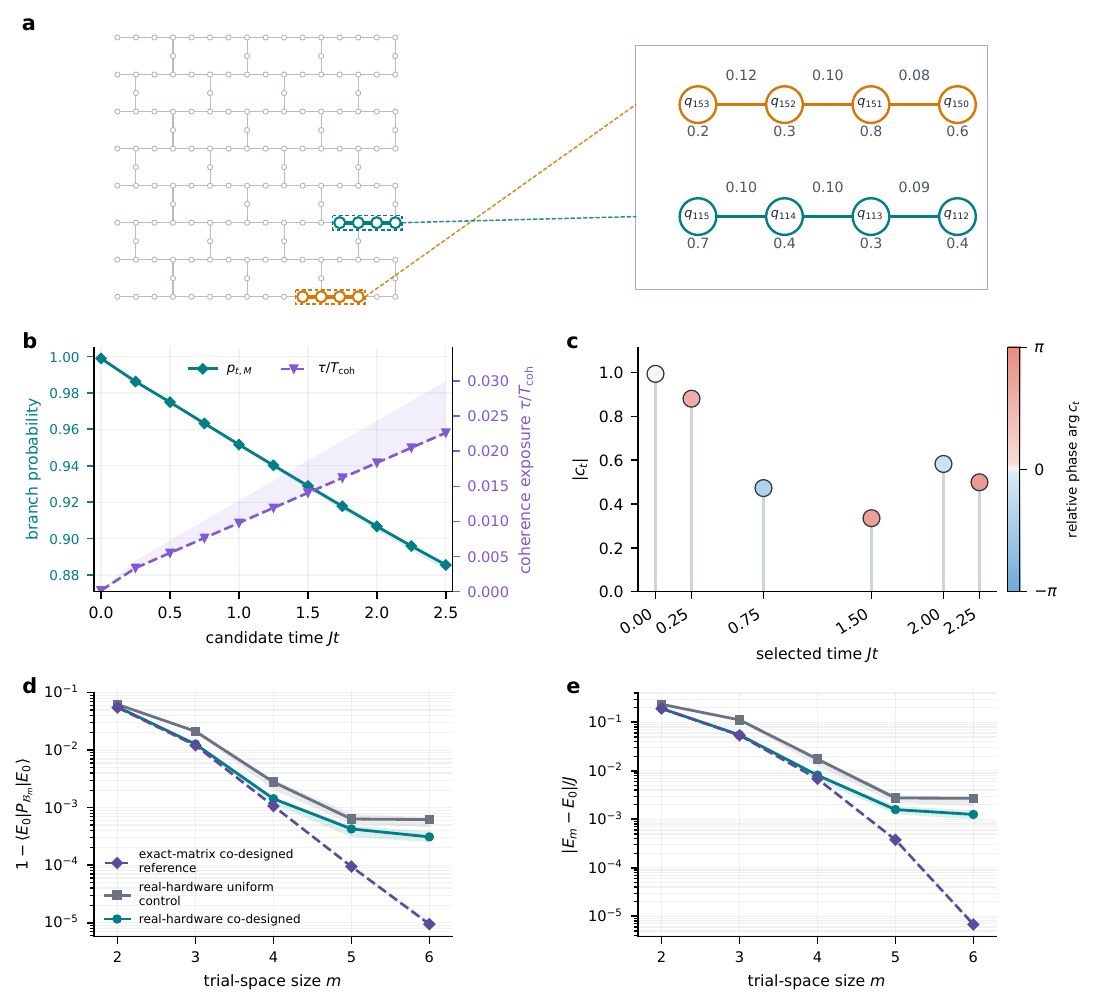}
\caption{\textbf{Platform-aware trial-space construction and measured four-qubit accuracy.}
\textbf{a,} Physical paths used during the seven analyzed acquisition days. Orange marks the path used on Day 1 and teal the path used on Days 2--7. Labels inside the circles are physical-qubit indices, numbers above the links are median (CZ)-error percentages, and numbers below the circles are median readout-error percentages.
\textbf{b,} Modeled response support and coherence exposure across the candidate-time dictionary.
\textbf{c,} Coefficients of the six-coordinate problem-informed trial space.
\textbf{d,e,} Missing target weight and projected-energy error for the exact-matrix reference, uniform-time control, and problem-informed policy. Curves and bands show cohort medians and interquartile ranges. The two hardware policies use the same physical path, shot budget, reconstruction, and projected-matrix analysis, so their difference reflects the selected time coordinates.}
\label{fig:ibm_kobe_codesign}
\end{figure*}

\begin{figure*}[t]
\centering
\includegraphics[width=0.98\textwidth]{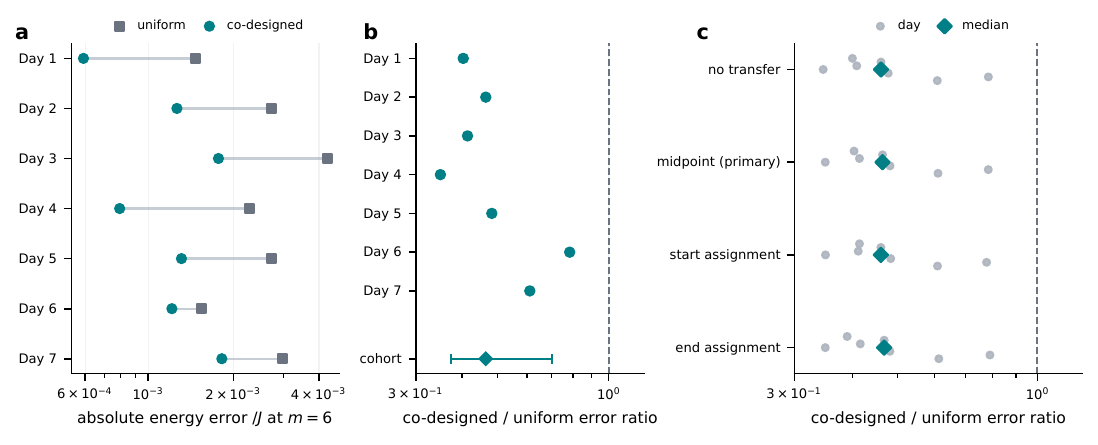}
\caption{\textbf{Problem-informed time coordinates improve projected-energy accuracy under real hardware noise.}
\textbf{a,} Paired \(m=6\) energy errors for the two policies on each of the seven analyzed acquisition days.
\textbf{b,} Corresponding ratios of problem-informed to uniform error. The cohort diamond shows the seven-day median ratio and paired-bootstrap 95\% interval; ratios below one favor the problem-informed policy.
\textbf{c,} Sensitivity to readout calibration and analysis-stage response transfer. Gray points denote individual days and diamonds their medians. The dashed line marks equal error.}
\label{fig:ibm_kobe_validation}
\end{figure*}

The archived-device study shows that platform information can participate in trial-space design before acquisition. We next test the resulting design logic on a live processor. The experiment keeps the physical problem, trial-space dimension, hardware path, shot budget, reconstruction, and projected-matrix analysis fixed, and compares two choices of evolution times: a problem-informed nonuniform set and a uniform control.

We use the same four-site open TFIM at \(h/J=0.9\), with input \(\lvert+\rangle^{\otimes4}\), an eleven-time candidate grid, and a six-coordinate budget. Before each acquisition, the problem-informed policy selects \(\{0,0.25,0.75,1.5,2.0,2.25\}\), while the control uses the uniform set \(\{0,0.5,1.0,1.5,2.0,2.5\}\). Both policies are executed on the same calibration-selected physical path on each date. A retrospective response-free selector chooses the same nonuniform times as the platform-constrained selector, so this experiment isolates the value of the problem-informed time policy under real acquisition noise rather than a calibration-induced change in the time set.

Figure~\ref{fig:ibm_kobe_codesign} summarizes the platform conditions and the resulting trial spaces. The selected physical path changes as the reported device profile changes (Figure~\ref{fig:ibm_kobe_codesign}a), while increasing evolution time lowers the modeled response support and increases coherence exposure (Figure~\ref{fig:ibm_kobe_codesign}b). Despite these changes in the platform profile, the same six nonuniform times are selected throughout the cohort. The exact projected calculation shows how those coordinates combine into the working trial space (Figure~\ref{fig:ibm_kobe_codesign}c).

The two policies are then measured through the same acquisition and reconstruction pipeline. As the trial-space dimension increases, the problem-informed coordinates retain more of the target state and give lower median projected-energy error than uniform sampling (Figure~\ref{fig:ibm_kobe_codesign}d,e). The exact-matrix reference shows the accuracy available from the chosen representation before measurement noise; its separation from the hardware curves quantifies the remaining finite-shot and acquisition errors.

The analysis reported here uses seven acquisitions on distinct days, labeled Day 1 through Day 7. At the final dimension \(m=6\), the problem-informed policy gives a smaller energy error on all seven days (Figure~\ref{fig:ibm_kobe_validation}a,b). The median ratio of problem-informed to uniform error is \(0.464\), with a paired-bootstrap 95\% interval of \([0.374,0.701]\). Thus the nonuniform policy reduces the typical projected-energy error by 54\% [30\%, 63\%] under otherwise matched hardware conditions. The result is also insensitive to using the start or end readout calibration and to removing the analysis-stage response transfer (Figure~\ref{fig:ibm_kobe_validation}c). Supplementary Section~6.2 and Table~S5 give the complete paired inference.

These controls place the observed improvement at the pre-acquisition choice of evolution times rather than at a downstream response correction. The live experiment therefore establishes a narrower result than the archived-device study: problem-informed time coordinates improve four-qubit projected simulation under real hardware noise, while the marginal contribution of calibration to the coordinate choice remains unresolved in this cohort.

A separate scale experiment tests how far the hardware-selection procedure can be deployed without claiming projected-energy accuracy. Circuits spanning systems up to \(N=30\) cover 135 of the 156 physical qubits, but the primary path-ranking comparison is not resolved. These measurements therefore establish execution coverage and deep-circuit survival, while the projected-energy accuracy result remains the four-qubit experiment above (Supplementary Section~6.3 and Fig.~S7).


\section{Discussion}

The main contribution of this work is a representation-design principle: the states and coordinates used in a projected quantum simulation need not be fixed before information about the physical response and the execution platform is considered.  A reproducible response can reshape the candidate states, while a platform profile can constrain which times and mappings remain practical. In both cases, the target Hamiltonian and physical objective remain unchanged; what is adapted is the representation used to approximate them.

The Hubbard study exposes both the opportunity and an important boundary of this idea. Under a matched selector, response-shaped candidates recover a much better low-energy span than closed evolution, showing that the gain cannot be explained by adaptive selection alone.  The same candidates, however, become strongly redundant: after stable overlap modes are retained, the exact-span advantage can disappear.  This distinction is central to the method. \emph{Representation compression and estimable compression are not the same problem.}  A useful response must therefore do more than reveal target-relevant directions; it must encode those directions in coordinates that remain distinguishable under finite-data estimation.

The additional controls show that this limitation is operational rather than purely numerical.  A straightforward projected-matrix estimator fails to recover the correct stable subspace even at very large sampling budgets, while explicit conditional-response circuits introduce enough additional gate cost to obscure the ideal representational benefit on the small topology studied here.  Scalable implementations will therefore require response models and coordinate choices designed together with stable estimators, for example through mode-aware measurement, structured matrix estimation, or alternative well-conditioned coordinates.

The transfer studies clarify what should and should not be expected to carry between physical systems.  The useful response is not universal: local charge fluctuations, spin defects, and molecular configuration structure lead to different response operators and different time scales.  What transfers is the design logic---use accessible physical structure to generate candidate directions, retain those that are both useful and viable, and solve the original Hamiltonian in the resulting space.  The molecular example is particularly useful in this respect because its response is derived from the Hamiltonian rather than from an assumed environment, showing that the same representation-design idea can extend beyond monitored open-system dynamics.

Introducing processor information completes the co-design picture.  Response shaping and platform screening are complementary stages rather than identical interventions: the first expands the candidate library, whereas the second determines which coordinates remain viable for acquisition.  In the archived-device study, the target problem and the platform profile jointly change the selected mapping and evolution times before noisy acquisition is introduced.  The live experiment tests a narrower but experimentally cleaner consequence.  Under matched hardware and acquisition conditions, a problem-informed nonuniform time policy reduces the typical four-qubit projected-energy error by 54\% [30--63\%] relative to uniform sampling across the seven analyzed days.  The response-free and platform-constrained selectors choose the same nonuniform times in this cohort, so the experiment validates the coordinate-selection stage under real hardware noise rather than a physical conditional-response implementation or a separate calibration-induced change in those coordinates.  Because this comparison spans seven days on one processor and one four-qubit target, it establishes repeatability within that setting rather than cross-device or asymptotic generality.

The present results do not yet establish scalable projected simulation. The larger-device experiment demonstrates execution coverage rather than a resolved energy-accuracy advantage, and scalable response estimation, well-conditioned projected-matrix acquisition, and low-overhead physical response preparation remain open problems.  These limitations also identify the next methodological target: platform-aware coordinates should ultimately be judged not only by the quality of the span they generate, but by the amount of target-relevant information that can be extracted from that span at realistic execution cost.

The broader implication is that representation choice can become part of the hardware--algorithm interface rather than an upstream decision made for an ideal machine.  Reproducible information about the physical problem and the platform can be used before acquisition to decide which states, times, and mappings are worth realizing.  The value of this approach is therefore not that noise becomes useful in general, but that structured response and device constraints provide additional information for designing the representation in which a quantum simulation is carried out.


\section*{Methods}

\subsection{Hubbard model and local-response construction}

The Hubbard calculations use the six-site periodic Hamiltonian
\begin{equation}
 H=-t\sum_{i=1}^{N}\sum_{\sigma=\uparrow,\downarrow}
 \left(
 c_{i\sigma}^{\dagger}c_{i+1,\sigma}
 +c_{i+1,\sigma}^{\dagger}c_{i\sigma}
 \right)
 +U\sum_{i=1}^{N}n_{i\uparrow}n_{i\downarrow}.
 \label{eq:hubbard_methods}
\end{equation}
We work in the half-filled sector \(N_\uparrow=N_\downarrow=3\), which has dimension 400, and set \(t=1\) and \(U/t=16\).  The input determinant has spin-up occupation on sites \(0,2,4\) and spin-down occupation on sites \(1,3,5\).  The exact ground-state energy, \(-1.0603936670402792\), is used only for benchmark diagnostics.

The local-response candidates use the operators in Eq.~\eqref{eq:local_hubbard_response}.  In the half-filled sector, \(R_{\rm loc}\) vanishes on configurations with one particle per site and assigns weight two to each doublon--hole pair.  Candidates are generated with
\begin{equation}
 H_{\rm eff}^{\rm loc}=H-\frac{i\mu}{2}R_{\rm loc},
\end{equation}
without diagonalizing the target Hamiltonian.  Because \(\lVert R_{\rm loc}\rVert=6\), we use \(\mu\in\{0.1,0.2,0.5,1,2\}\), chosen so that the fastest amplitude-decay time \(2/(\mu\lVert R_{\rm loc}\rVert)\) spans the range from the full evolution window to approximately one time interval.  All matched comparisons use the same input state, candidate-time support, accepted dimension, and overlap treatment.

Exact eigenvectors enter only the reported energy, fidelity, and target-projection diagnostics.  An eigensystem-derived spectral response is used solely as a reference benchmark for the local construction; its definition and interaction-strength sweep are given in Supplementary Section~2.

\subsection{Conditional-response preparation}

For \(q_i=L_i^2\), each \(q_i\) projects onto an empty or doubly occupied site.  Since these projectors commute,
\begin{equation}
 e^{-\mu\Delta t R_{\rm loc}/2}
 =
 \prod_i
 \left[
 (I-q_i)+e^{-\mu\Delta t/2}q_i
 \right].
 \label{eq:conditional_factorization}
\end{equation}
Each factor can be implemented by coupling an ancilla conditionally on \(q_i=1\) and retaining the zero-outcome branch.  A dynamic implementation may measure, reset, and reuse one ancilla between factors; alternatively, a static implementation may use separate ancillas, defer their measurements, and postselect the all-zero record.  The latter avoids mid-circuit reset at the cost of additional ancillas.

The finite-step implementation uses the symmetric sequence
\begin{equation}
 e^{-iH\Delta t/2}
 e^{-\mu\Delta tR_{\rm loc}/2}
 e^{-iH\Delta t/2},
\end{equation}
whose local splitting error is \(O(\Delta t^3)\) and accumulated error is \(O(t\Delta t^2)\).  The probability of preparing a retained candidate is the squared norm of its unnormalized conditional branch.  Failed postselections are therefore included in the attempted-execution cost. Finite-step, thermal, and finite-data controls are reported in Supplementary Section~3 and Fig.~S2.

\subsection{Trial-space selection and cross-system design}

For the Hubbard study, selection begins from the common \(t=0\) state and greedily combines four quantities: novelty relative to the current span, normalized alignment with the current Ritz residual, a branch-derived support factor, and an overlap-conditioning penalty.  Clean representation errors are evaluated after orthonormalizing the selected span, whereas \(\kappa(S)\) refers to the original nonorthogonal coordinates.

The selector-matched closed-time control uses the same novelty, residual, conditioning, stopping, and tie rules as the local-response construction. The prescribed closed-time baseline in Fig.~\ref{fig:hubbard_representation} instead uses a fixed nested time order. Eigensystem-derived-response, fixed-response, and survival-matched controls are described in Supplementary Section~2.

The transfer study in Fig.~\ref{fig:cross_model_representation} selects a complete response strength and time window for each physical problem rather than reusing the Hubbard selector weights.  Every design contains twelve uniformly spaced coordinates within its trial window.  The standard TFIM windows are
\[
T\in\{4,8,12,16,20,24\},
\]
with additional windows
\[
T\in\{48,96,128,192\}
\]
for \(h/J=0.5\).  The N$_2$ windows are
\[
T\in\{4,8,12,16,20,24,32,40,48\}.
\]
The TFIM response strengths are
\[
\mu\in\{0.02,0.05,0.1,0.2,0.5,1,2,5\},
\]
and the molecular filter strengths are
\[
\eta\in\{0.005,0.01,0.02,0.05,0.1,0.2,0.5,1,2,5,10,20\}.
\]

A TFIM design is retained only when its minimum conditional-branch probability is at least \(10^{-3}\) and its raw overlap condition number does not exceed \(10^8\).  The molecular construction applies the analogous norm-retention and conditioning requirements as algorithmic support criteria rather than interpreting the filter norm as a measured decay probability.  Among feasible designs, the one with the lowest projected Ritz energy is selected.  The exact target eigenvalue or eigenvector is not used in this optimization.

Exact candidate states, \(H\)-actions, Gram matrices, residuals, and projected quantities serve as controlled proxies in these classically tractable tests.  A scalable implementation would estimate them or replace them with measurable surrogates.  Exact diagonalization is used only for benchmark energies and fidelities and for the explicitly labeled reference calculations.  For Fig.~\ref{fig:cross_model_representation}, the evaluation target is the lowest exact eigenstate whose squared overlap with the common input exceeds \(10^{-12}\).  This is the absolute ground state except for stretched N$_2$, where the rule selects eigenstate index three.

\subsection{Archived-device response model and simulated acquisition}

The archived-device study uses the configuration and calibration record for \texttt{ibm\_manila} dated May 27, 2024, as distributed with \texttt{qiskit-ibm-runtime} 0.47.0.  The archived configuration and properties files, normalized SI-unit data, source URLs, package versions, dates, and SHA-256 hashes are retained with the source data.

The platform profile provides connectivity, native operations and durations, \(T_1\), \(T_2\), gate errors, and asymmetric readout errors.  For qubit \(q\), relaxation and pure-dephasing rates are represented by
\begin{equation}
 \Gamma_{1,q}=T_{1,q}^{-1},
 \quad
 \Gamma_{\phi,q}
 =T_{2,q}^{-1}-\frac{1}{2T_{1,q}} .
 \label{eq:device_rates}
\end{equation}
Candidate circuits are transpiled with first-order product-formula steps \(J\Delta t=0.25\), optimization level 1, deterministic seed 271828, and the archived coupling graph and native basis.  Native-operation dependencies define an ASAP schedule, from which active and idle exposure are obtained.

Amplitude relaxation contributes a state-dependent no-amplitude-jump factor along the compiled schedule.  Pure dephasing contributes scalar survival. For each native gate, a residual depolarizing contribution is chosen so that thermal and residual infidelity reproduce the archived gate error; if the thermal contribution already exceeds that value, the residual is set to zero and the record is flagged.  Correlated errors, leakage, crosstalk, non-Markovian effects, and drift that are absent from the archive are not imputed.  Readout error is treated separately through the archived asymmetric assignment matrices.  The complete operational-response model is given in Supplementary Section~5.

For the four-site TFIM, candidate coordinates consist of eleven times \(Jt=0,0.25,\ldots,2.5\) and all \(5P4=120\) ordered four-qubit mappings. The platform-aware search considers six retained-coordinate windows on this grid.  A design is discarded if compilation leaves the selected subgraph, if its minimum modeled support falls below \(10^{-3}\), if \(\kappa(S)>10^8\), or if any candidate schedule exceeds \(T_{\rm coh}(M)=\min_{q\in M}\{T_{1,q},T_{2,q}\}\).  The feasible design with the lowest projected Ritz energy is retained without using a target eigenpair.

Two controls separate representation design from the complete noisy workflow.  The matched response-free sweep uses the same candidate dictionary, admissibility conditions, and projected objective but removes the device-response reshaping during candidate construction.  A separate conventional acquisition baseline fixes its mapping and uniform time coordinates before the calibration record is read.  The former isolates the effect of platform information on representation design; the latter defines the end-to-end noisy-workflow comparison.

For simulated matrix acquisition, each of the six selected precursor states is measured in all \(3^4\) local Pauli settings.  Circuits are transpiled at optimization level 2 and propagated through the archived gate, idle, and readout noise model.  Each setting uses 16,384 shots in each of six fixed batches.  Assignment correction is applied setting by setting, compatible measurements are averaged to obtain the required Pauli expectations, and linear inversion is projected onto the positive trace-one cone.

The closed precursor has exact global-\(X\) symmetry.  After verifying this symmetry, its principal reconstructed vector is extracted.  For the platform-aware reconstruction, the ideal compiled sequence is inverted and the same schedule is replayed with the no-amplitude-jump response inserted after active and idle intervals.  The resulting vectors are used consistently to construct both the overlap and Hamiltonian matrices.  These matrices are Hermitized and solved after a relative overlap-eigenvalue cutoff of \(10^{-5}\).  Directly measured \(X_i\) and \(Z_iZ_{i+1}\) expectation values are retained as diagnostics but are not mixed with the reconstructed off-diagonal projected matrix elements.

The conventional noisy baseline uses the same hardware noise model, local measurement settings, shot budget, positive trace-one reconstruction, and overlap cutoff.  Its mapping and times are fixed before calibration is read, and it omits the calibration-derived assignment correction, precursor-symmetry verification, and response transfer.  It therefore defines an end-to-end workflow comparison rather than an isolated test of trial-space design.  The adopted shallow estimator, conditional-preparation cost control, and reconstruction diagnostics are reported in Supplementary Section~5.1.

Archived extensions to larger processor graphs are given in Supplementary Sections~5.2--5.3.  These tests use the same design logic but do not replace the four-qubit projected-matrix acquisition used for the main controlled comparison.

\subsection{Live-hardware acquisition and paired inference}

The live experiment was executed on the 156-qubit Heron processor \texttt{ibm\_kobe} using Qiskit 2.4.1 and \texttt{qiskit-ibm-runtime} 0.47.0.  Each returned acquisition directory contains the source and transpiled circuits, resource manifests, backend configuration and property snapshots, Runtime identifiers and timestamps, primitive metadata, counts, shot-level bitstrings, and nested SHA-256 manifests.  All stored count histograms were independently reconstructed from the returned bitstrings.

The primary projected-energy tier uses the four-site open TFIM at \(h/J=0.9\), input \(\lvert+\rangle^{\otimes4}\), candidate grid \(Jt=0,0.25,\ldots,2.5\), and retained dimension \(m=6\).  Before each science acquisition, the problem-informed policy evaluates all 252 six-coordinate subsets containing \(Jt=0\), using the target Hamiltonian and the contemporaneous platform record but no target eigenpair.  The active control uses the fixed uniform coordinates \(\{0,0.5,1.0,1.5,2.0,2.5\}\) and a calibration-only path score.  A retrospective response-free selector is evaluated separately to determine whether the platform model changes the nonuniform coordinate set itself.

Shared precursor circuits are acquired once and reused by both policies. Exact deduplication gives 729 science circuits per acquisition block.  Each precursor is measured in all \(3^4\) local Pauli settings with 4096 shots. Sixteen computational-basis assignment circuits are acquired immediately before and after the science phase, giving 761 circuits and 3,117,056 shots per block.  Each tier is executed within a single exclusive Runtime Session in the fixed order
\[
\text{start controls}\rightarrow\text{science}\rightarrow\text{end controls}.
\]
A complete bracketed acquisition on a distinct UTC day is the independent statistical unit.  Provider timestamps and calibration fingerprints are used for provenance rather than as independent observations.  The analysis reported here uses seven acquisitions on distinct days.

For each Pauli setting, the primary readout correction uses the arithmetic midpoint of the empirical start and end \(16\times16\) assignment matrices. Both policies then use the same constrained assignment correction, positive trace-one reconstruction, precursor-symmetry verification, projected overlap and Hamiltonian construction, and relative \(10^{-5}\) overlap cutoff. Start-only, end-only, and symmetry-closed reconstructions provide sensitivity analyses.

For acquisition date \(d\), the primary contrast is
\begin{equation}
 \Delta_d=
 \log_{10}
 \left(
 \frac{\epsilon_{{\rm task},d}}
      {\epsilon_{{\rm ctrl},d}}
 \right)
 \qquad (m=6).
 \label{eq:hardware_log_ratio}
\end{equation}
The cohort statistic is the median of \(\Delta_d\) over the seven analyzed days. Within each day, 4096 complete multinomial circuit/shot resamples preserve the shared-circuit structure and resample both assignment brackets before reconstructing the projected problem.  A second stage resamples the seven day blocks 20,000 times.  All bootstrap calculations use seed 20260806. Improvement is declared when the upper 97.5th percentile of the paired-bootstrap interval lies below zero and the valid-solution fraction is at least 0.99 at both resampling stages.  Exact sign tests, leave-one-day-out summaries, alternative readout corrections, and the remaining secondary paired sensitivity analysis are reported in Supplementary Section~6.2.

A separate scale tier evaluates \(N=\{4,8,16,24,30\}\) on three calibration-ranked path strata: calibration-best, upper-quartile, and beam-median.  Four shallow evolution times through \(Jt=2.5\) are measured with 2048 shots per circuit.  Each of the five analyzed date blocks contains 150 shallow science circuits, 12 symmetric-second-order proxy-depth sentinels, and 90 bracket controls, for 252 circuits per block.  The primary endpoint is the across-date summary of the smaller energy advantage of the calibration-best path over the two comparator strata.  Its bootstrap uses 5000 control-and-science resamples and the five-date denominator.  The four individual energy- and mirror-return contrasts are treated as multiplicity-unadjusted secondary summaries.

The proxy-depth sentinel measures forward energy and local and global mirror return at the schedule endpoints for \(N=8,16,24,30\).  Because this tier does not reconstruct a projected eigensolve or provide an ideal deep-circuit reference, it is interpreted as an execution-coverage and survival test rather than a large-system projected-energy measurement.  Full acquisition integrity and reconstruction details are given in Supplementary Sections~6.1--6.2 and Table~S5; active resource accounting is given in Table~S6; scale-tier sensitivities are given in Supplementary Section~6.3 and Fig.~S7.

\section*{Data availability}

Source data underlying the numerical and real-hardware main and Supplementary
figures and tables are provided with this paper in the accompanying Source
Data archive. For the live-hardware campaign, this archive contains the analyzer
outputs, processed analysis tables, panel mappings, statistical units, random
seeds, exact resource accounting, and SHA-256 links to the acquisition folders
used in the reported seven-day analysis. The complete folders, including raw counts, shot-level
bitstrings, primitive metadata, Runtime receipts, backend snapshots,
compiled-circuit and resource manifests, and versioned protocol and result
records, are supplied in the review package and will be archived in a
persistent public repository before publication.

\section*{Code availability}

Versioned code for preparing, submitting, retrieving, verifying, and analyzing
the Kobe experiment, together with code for reproducing the simulations,
source-data tables, figures, and tests, is supplied in the review package. A
DOI-minting public archive and software license will be added before
publication.

\section*{Author contributions}

B.P. conceived the structured-response channel-orbit framework, developed the
prototypes, wrote the original draft, and supervised the study. H.P. executed
the experiments on \texttt{ibm\_kobe}. B.P., V.A., and H.P. developed the
numerical workflow, analyzed the results, and prepared the manuscript.

\section*{Competing interests}

The authors declare no competing interests.

\section*{Acknowledgements}
This work was supported by the Early Career Research Program of the U.S. Department of Energy, Office of Science, under Grant No. FWP 83466. The contribution of H.P. to this paper is based on results obtained from a project, JPNP20017, commissioned by the New Energy and Industrial Technology Development Organization (NEDO). H.P. is also supported by the RIKEN TRIP initiative (RIKEN Quantum).

\bibliographystyle{apsrev4-2}
\bibliography{ref}

\begin{thebibliography}{38}%
\makeatletter
\providecommand \@ifxundefined [1]{%
 \@ifx{#1\undefined}
}%
\providecommand \@ifnum [1]{%
 \ifnum #1\expandafter \@firstoftwo
 \else \expandafter \@secondoftwo
 \fi
}%
\providecommand \@ifx [1]{%
 \ifx #1\expandafter \@firstoftwo
 \else \expandafter \@secondoftwo
 \fi
}%
\providecommand \natexlab [1]{#1}%
\providecommand \enquote  [1]{``#1''}%
\providecommand \bibnamefont  [1]{#1}%
\providecommand \bibfnamefont [1]{#1}%
\providecommand \citenamefont [1]{#1}%
\providecommand \href@noop [0]{\@secondoftwo}%
\providecommand \href [0]{\begingroup \@sanitize@url \@href}%
\providecommand \@href[1]{\@@startlink{#1}\@@href}%
\providecommand \@@href[1]{\endgroup#1\@@endlink}%
\providecommand \@sanitize@url [0]{\catcode `\\12\catcode `\$12\catcode
  `\&12\catcode `\#12\catcode `\^12\catcode `\_12\catcode `\%12\relax}%
\providecommand \@@startlink[1]{}%
\providecommand \@@endlink[0]{}%
\providecommand \url  [0]{\begingroup\@sanitize@url \@url }%
\providecommand \@url [1]{\endgroup\@href {#1}{\urlprefix }}%
\providecommand \urlprefix  [0]{URL }%
\providecommand \Eprint [0]{\href }%
\providecommand \doibase [0]{https://doi.org/}%
\providecommand \selectlanguage [0]{\@gobble}%
\providecommand \bibinfo  [0]{\@secondoftwo}%
\providecommand \bibfield  [0]{\@secondoftwo}%
\providecommand \translation [1]{[#1]}%
\providecommand \BibitemOpen [0]{}%
\providecommand \bibitemStop [0]{}%
\providecommand \bibitemNoStop [0]{.\EOS\space}%
\providecommand \EOS [0]{\spacefactor3000\relax}%
\providecommand \BibitemShut  [1]{\csname bibitem#1\endcsname}%
\let\auto@bib@innerbib\@empty
\bibitem [{\citenamefont {Alexeev}\ \emph {et~al.}(2021)\citenamefont {Alexeev}
  \emph {et~al.}}]{Alexeev2021CoDesign}%
  \BibitemOpen
  \bibfield  {author} {\bibinfo {author} {\bibfnamefont {Y.}~\bibnamefont
  {Alexeev}} \emph {et~al.},\ }\href
  {https://doi.org/10.1103/PRXQuantum.2.017001} {\bibfield  {journal} {\bibinfo
   {journal} {PRX Quantum}\ }\textbf {\bibinfo {volume} {2}},\ \bibinfo {pages}
  {017001} (\bibinfo {year} {2021})}\BibitemShut {NoStop}%
\bibitem [{\citenamefont {Wittler}\ \emph {et~al.}(2021)\citenamefont
  {Wittler}, \citenamefont {Roy}, \citenamefont {Pack}, \citenamefont
  {Werninghaus}, \citenamefont {Roy}, \citenamefont {Egger}, \citenamefont
  {Filipp}, \citenamefont {Wilhelm},\ and\ \citenamefont
  {Machnes}}]{Wittler2021C3}%
  \BibitemOpen
  \bibfield  {author} {\bibinfo {author} {\bibfnamefont {N.}~\bibnamefont
  {Wittler}}, \bibinfo {author} {\bibfnamefont {F.}~\bibnamefont {Roy}},
  \bibinfo {author} {\bibfnamefont {K.}~\bibnamefont {Pack}}, \bibinfo {author}
  {\bibfnamefont {M.}~\bibnamefont {Werninghaus}}, \bibinfo {author}
  {\bibfnamefont {A.~S.}\ \bibnamefont {Roy}}, \bibinfo {author} {\bibfnamefont
  {D.~J.}\ \bibnamefont {Egger}}, \bibinfo {author} {\bibfnamefont
  {S.}~\bibnamefont {Filipp}}, \bibinfo {author} {\bibfnamefont {F.~K.}\
  \bibnamefont {Wilhelm}},\ and\ \bibinfo {author} {\bibfnamefont
  {S.}~\bibnamefont {Machnes}},\ }\href
  {https://doi.org/10.1103/PhysRevApplied.15.034080} {\bibfield  {journal}
  {\bibinfo  {journal} {Physical Review Applied}\ }\textbf {\bibinfo {volume}
  {15}},\ \bibinfo {pages} {034080} (\bibinfo {year} {2021})},\ \Eprint
  {https://arxiv.org/abs/2009.09866} {arXiv:2009.09866} \BibitemShut {NoStop}%
\bibitem [{\citenamefont {Javadi-Abhari}\ \emph {et~al.}(2024)\citenamefont
  {Javadi-Abhari}, \citenamefont {Treinish}, \citenamefont {Krsulich},
  \citenamefont {Wood}, \citenamefont {Lishman}, \citenamefont {Gacon},
  \citenamefont {Martiel}, \citenamefont {Nation}, \citenamefont {Bishop},
  \citenamefont {Cross}, \citenamefont {Johnson},\ and\ \citenamefont
  {Gambetta}}]{JavadiAbhari2024Qiskit}%
  \BibitemOpen
  \bibfield  {author} {\bibinfo {author} {\bibfnamefont {A.}~\bibnamefont
  {Javadi-Abhari}}, \bibinfo {author} {\bibfnamefont {M.}~\bibnamefont
  {Treinish}}, \bibinfo {author} {\bibfnamefont {K.}~\bibnamefont {Krsulich}},
  \bibinfo {author} {\bibfnamefont {C.~J.}\ \bibnamefont {Wood}}, \bibinfo
  {author} {\bibfnamefont {J.}~\bibnamefont {Lishman}}, \bibinfo {author}
  {\bibfnamefont {J.}~\bibnamefont {Gacon}}, \bibinfo {author} {\bibfnamefont
  {S.}~\bibnamefont {Martiel}}, \bibinfo {author} {\bibfnamefont {P.~D.}\
  \bibnamefont {Nation}}, \bibinfo {author} {\bibfnamefont {L.~S.}\
  \bibnamefont {Bishop}}, \bibinfo {author} {\bibfnamefont {A.~W.}\
  \bibnamefont {Cross}}, \bibinfo {author} {\bibfnamefont {B.~R.}\ \bibnamefont
  {Johnson}},\ and\ \bibinfo {author} {\bibfnamefont {J.~M.}\ \bibnamefont
  {Gambetta}},\ }\bibfield  {journal} {\bibinfo  {journal} {arXiv preprint
  arXiv:2405.08810}\ }\href {https://doi.org/10.48550/arXiv.2405.08810}
  {10.48550/arXiv.2405.08810} (\bibinfo {year} {2024})\BibitemShut {NoStop}%
\bibitem [{\citenamefont {Miessen}\ \emph {et~al.}(2023)\citenamefont
  {Miessen}, \citenamefont {Ollitrault}, \citenamefont {Tacchino},\ and\
  \citenamefont {Tavernelli}}]{Miessen2023QuantumDynamics}%
  \BibitemOpen
  \bibfield  {author} {\bibinfo {author} {\bibfnamefont {A.}~\bibnamefont
  {Miessen}}, \bibinfo {author} {\bibfnamefont {P.~J.}\ \bibnamefont
  {Ollitrault}}, \bibinfo {author} {\bibfnamefont {F.}~\bibnamefont
  {Tacchino}},\ and\ \bibinfo {author} {\bibfnamefont {I.}~\bibnamefont
  {Tavernelli}},\ }\href {https://doi.org/10.1038/s43588-022-00374-2}
  {\bibfield  {journal} {\bibinfo  {journal} {Nature Computational Science}\
  }\textbf {\bibinfo {volume} {3}},\ \bibinfo {pages} {25} (\bibinfo {year}
  {2023})}\BibitemShut {NoStop}%
\bibitem [{\citenamefont {Fauseweh}(2024)}]{Fauseweh2024ManyBody}%
  \BibitemOpen
  \bibfield  {author} {\bibinfo {author} {\bibfnamefont {B.}~\bibnamefont
  {Fauseweh}},\ }\href {https://doi.org/10.1038/s41467-024-46402-9} {\bibfield
  {journal} {\bibinfo  {journal} {Nature Communications}\ }\textbf {\bibinfo
  {volume} {15}},\ \bibinfo {pages} {2123} (\bibinfo {year}
  {2024})}\BibitemShut {NoStop}%
\bibitem [{\citenamefont {Battaglia}\ \emph {et~al.}(2024)\citenamefont
  {Battaglia}, \citenamefont {Rossmannek}, \citenamefont {Rybkin},
  \citenamefont {Tavernelli},\ and\ \citenamefont
  {Hutter}}]{Battaglia2024ActiveSpace}%
  \BibitemOpen
  \bibfield  {author} {\bibinfo {author} {\bibfnamefont {S.}~\bibnamefont
  {Battaglia}}, \bibinfo {author} {\bibfnamefont {M.}~\bibnamefont
  {Rossmannek}}, \bibinfo {author} {\bibfnamefont {V.~V.}\ \bibnamefont
  {Rybkin}}, \bibinfo {author} {\bibfnamefont {I.}~\bibnamefont {Tavernelli}},\
  and\ \bibinfo {author} {\bibfnamefont {J.}~\bibnamefont {Hutter}},\ }\href
  {https://doi.org/10.1038/s41524-024-01477-2} {\bibfield  {journal} {\bibinfo
  {journal} {npj Computational Materials}\ }\textbf {\bibinfo {volume} {10}},\
  \bibinfo {pages} {297} (\bibinfo {year} {2024})}\BibitemShut {NoStop}%
\bibitem [{\citenamefont {Pardo}\ \emph {et~al.}(2023)\citenamefont {Pardo},
  \citenamefont {Greenberg}, \citenamefont {Fortinsky}, \citenamefont {Katz},\
  and\ \citenamefont {Zohar}}]{Pardo2023Gauge}%
  \BibitemOpen
  \bibfield  {author} {\bibinfo {author} {\bibfnamefont {G.}~\bibnamefont
  {Pardo}}, \bibinfo {author} {\bibfnamefont {T.}~\bibnamefont {Greenberg}},
  \bibinfo {author} {\bibfnamefont {A.}~\bibnamefont {Fortinsky}}, \bibinfo
  {author} {\bibfnamefont {N.}~\bibnamefont {Katz}},\ and\ \bibinfo {author}
  {\bibfnamefont {E.}~\bibnamefont {Zohar}},\ }\href
  {https://doi.org/10.1103/PhysRevResearch.5.023077} {\bibfield  {journal}
  {\bibinfo  {journal} {Physical Review Research}\ }\textbf {\bibinfo {volume}
  {5}},\ \bibinfo {pages} {023077} (\bibinfo {year} {2023})}\BibitemShut
  {NoStop}%
\bibitem [{\citenamefont {Weimer}\ \emph {et~al.}(2021)\citenamefont {Weimer},
  \citenamefont {Kshetrimayum},\ and\ \citenamefont
  {Or{\'u}s}}]{Weimer2021OpenSystems}%
  \BibitemOpen
  \bibfield  {author} {\bibinfo {author} {\bibfnamefont {H.}~\bibnamefont
  {Weimer}}, \bibinfo {author} {\bibfnamefont {A.}~\bibnamefont
  {Kshetrimayum}},\ and\ \bibinfo {author} {\bibfnamefont {R.}~\bibnamefont
  {Or{\'u}s}},\ }\href {https://doi.org/10.1103/RevModPhys.93.015008}
  {\bibfield  {journal} {\bibinfo  {journal} {Reviews of Modern Physics}\
  }\textbf {\bibinfo {volume} {93}},\ \bibinfo {pages} {015008} (\bibinfo
  {year} {2021})}\BibitemShut {NoStop}%
\bibitem [{\citenamefont {McClean}\ \emph {et~al.}(2017)\citenamefont
  {McClean}, \citenamefont {Schwartz}, \citenamefont {Carter},\ and\
  \citenamefont {de~Jong}}]{McClean2017QSE}%
  \BibitemOpen
  \bibfield  {author} {\bibinfo {author} {\bibfnamefont {J.~R.}\ \bibnamefont
  {McClean}}, \bibinfo {author} {\bibfnamefont {M.~E.}\ \bibnamefont
  {Schwartz}}, \bibinfo {author} {\bibfnamefont {J.}~\bibnamefont {Carter}},\
  and\ \bibinfo {author} {\bibfnamefont {W.~A.}\ \bibnamefont {de~Jong}},\
  }\href {https://doi.org/10.1103/PhysRevA.95.042308} {\bibfield  {journal}
  {\bibinfo  {journal} {Physical Review A}\ }\textbf {\bibinfo {volume} {95}},\
  \bibinfo {pages} {042308} (\bibinfo {year} {2017})}\BibitemShut {NoStop}%
\bibitem [{\citenamefont {Epperly}\ \emph {et~al.}(2022)\citenamefont
  {Epperly}, \citenamefont {Lin},\ and\ \citenamefont
  {Nakatsukasa}}]{Epperly2022QSD}%
  \BibitemOpen
  \bibfield  {author} {\bibinfo {author} {\bibfnamefont {E.~N.}\ \bibnamefont
  {Epperly}}, \bibinfo {author} {\bibfnamefont {L.}~\bibnamefont {Lin}},\ and\
  \bibinfo {author} {\bibfnamefont {Y.}~\bibnamefont {Nakatsukasa}},\ }\href
  {https://doi.org/10.1137/21M145954X} {\bibfield  {journal} {\bibinfo
  {journal} {SIAM Journal on Matrix Analysis and Applications}\ }\textbf
  {\bibinfo {volume} {43}},\ \bibinfo {pages} {1263} (\bibinfo {year}
  {2022})}\BibitemShut {NoStop}%
\bibitem [{\citenamefont {Shen}\ \emph {et~al.}(2023)\citenamefont {Shen},
  \citenamefont {Klymko}, \citenamefont {Sud}, \citenamefont {Williams-Young},
  \citenamefont {de~Jong},\ and\ \citenamefont {Tubman}}]{Shen2023RTK}%
  \BibitemOpen
  \bibfield  {author} {\bibinfo {author} {\bibfnamefont {Y.}~\bibnamefont
  {Shen}}, \bibinfo {author} {\bibfnamefont {K.}~\bibnamefont {Klymko}},
  \bibinfo {author} {\bibfnamefont {J.}~\bibnamefont {Sud}}, \bibinfo {author}
  {\bibfnamefont {D.~B.}\ \bibnamefont {Williams-Young}}, \bibinfo {author}
  {\bibfnamefont {W.~A.}\ \bibnamefont {de~Jong}},\ and\ \bibinfo {author}
  {\bibfnamefont {N.~M.}\ \bibnamefont {Tubman}},\ }\href
  {https://doi.org/10.22331/q-2023-07-25-1066} {\bibfield  {journal} {\bibinfo
  {journal} {Quantum}\ }\textbf {\bibinfo {volume} {7}},\ \bibinfo {pages}
  {1066} (\bibinfo {year} {2023})}\BibitemShut {NoStop}%
\bibitem [{\citenamefont {Parrish}\ and\ \citenamefont
  {McMahon}(2019)}]{Parrish2019QFD}%
  \BibitemOpen
  \bibfield  {author} {\bibinfo {author} {\bibfnamefont {R.~M.}\ \bibnamefont
  {Parrish}}\ and\ \bibinfo {author} {\bibfnamefont {P.~L.}\ \bibnamefont
  {McMahon}},\ }\href {https://doi.org/10.48550/arXiv.1909.08925} {\bibinfo
  {title} {Quantum filter diagonalization: Quantum eigendecomposition without
  full quantum phase estimation}} (\bibinfo {year} {2019}),\ \Eprint
  {https://arxiv.org/abs/1909.08925} {arXiv:1909.08925 [quant-ph]} \BibitemShut
  {NoStop}%
\bibitem [{\citenamefont {Cohn}\ \emph {et~al.}(2021)\citenamefont {Cohn},
  \citenamefont {Motta},\ and\ \citenamefont {Parrish}}]{Cohn2021QFD}%
  \BibitemOpen
  \bibfield  {author} {\bibinfo {author} {\bibfnamefont {J.}~\bibnamefont
  {Cohn}}, \bibinfo {author} {\bibfnamefont {M.}~\bibnamefont {Motta}},\ and\
  \bibinfo {author} {\bibfnamefont {R.~M.}\ \bibnamefont {Parrish}},\ }\href
  {https://doi.org/10.1103/PRXQuantum.2.040352} {\bibfield  {journal} {\bibinfo
   {journal} {PRX Quantum}\ }\textbf {\bibinfo {volume} {2}},\ \bibinfo {pages}
  {040352} (\bibinfo {year} {2021})}\BibitemShut {NoStop}%
\bibitem [{\citenamefont {Motta}\ \emph {et~al.}(2020)\citenamefont {Motta},
  \citenamefont {Sun}, \citenamefont {Tan}, \citenamefont {O'Rourke},
  \citenamefont {Ye}, \citenamefont {Minnich}, \citenamefont {Brand{\~a}o},\
  and\ \citenamefont {Chan}}]{Motta2020QITE}%
  \BibitemOpen
  \bibfield  {author} {\bibinfo {author} {\bibfnamefont {M.}~\bibnamefont
  {Motta}}, \bibinfo {author} {\bibfnamefont {C.}~\bibnamefont {Sun}}, \bibinfo
  {author} {\bibfnamefont {A.~T.~K.}\ \bibnamefont {Tan}}, \bibinfo {author}
  {\bibfnamefont {M.~J.}\ \bibnamefont {O'Rourke}}, \bibinfo {author}
  {\bibfnamefont {E.}~\bibnamefont {Ye}}, \bibinfo {author} {\bibfnamefont
  {A.~J.}\ \bibnamefont {Minnich}}, \bibinfo {author} {\bibfnamefont {F.~G.
  S.~L.}\ \bibnamefont {Brand{\~a}o}},\ and\ \bibinfo {author} {\bibfnamefont
  {G.~K.-L.}\ \bibnamefont {Chan}},\ }\href
  {https://doi.org/10.1038/s41567-019-0704-4} {\bibfield  {journal} {\bibinfo
  {journal} {Nature Physics}\ }\textbf {\bibinfo {volume} {16}},\ \bibinfo
  {pages} {205} (\bibinfo {year} {2020})}\BibitemShut {NoStop}%
\bibitem [{\citenamefont {Yeter-Aydeniz}\ \emph {et~al.}(2020)\citenamefont
  {Yeter-Aydeniz}, \citenamefont {Pooser},\ and\ \citenamefont
  {Siopsis}}]{YeterAydeniz2020QLanczos}%
  \BibitemOpen
  \bibfield  {author} {\bibinfo {author} {\bibfnamefont {K.}~\bibnamefont
  {Yeter-Aydeniz}}, \bibinfo {author} {\bibfnamefont {R.~C.}\ \bibnamefont
  {Pooser}},\ and\ \bibinfo {author} {\bibfnamefont {G.}~\bibnamefont
  {Siopsis}},\ }\href {https://doi.org/10.1038/s41534-020-00290-1} {\bibfield
  {journal} {\bibinfo  {journal} {npj Quantum Information}\ }\textbf {\bibinfo
  {volume} {6}},\ \bibinfo {pages} {63} (\bibinfo {year} {2020})}\BibitemShut
  {NoStop}%
\bibitem [{\citenamefont {Jones}\ \emph {et~al.}(2022)\citenamefont {Jones},
  \citenamefont {Vallury}, \citenamefont {Hill},\ and\ \citenamefont
  {Hollenberg}}]{Jones2022QCM}%
  \BibitemOpen
  \bibfield  {author} {\bibinfo {author} {\bibfnamefont {M.~A.}\ \bibnamefont
  {Jones}}, \bibinfo {author} {\bibfnamefont {H.~J.}\ \bibnamefont {Vallury}},
  \bibinfo {author} {\bibfnamefont {C.~D.}\ \bibnamefont {Hill}},\ and\
  \bibinfo {author} {\bibfnamefont {L.~C.~L.}\ \bibnamefont {Hollenberg}},\
  }\href {https://doi.org/10.1038/s41598-022-12324-z} {\bibfield  {journal}
  {\bibinfo  {journal} {Scientific Reports}\ }\textbf {\bibinfo {volume}
  {12}},\ \bibinfo {pages} {8985} (\bibinfo {year} {2022})}\BibitemShut
  {NoStop}%
\bibitem [{\citenamefont {Klymko}\ \emph {et~al.}(2022)\citenamefont {Klymko},
  \citenamefont {Mejuto-Zaera}, \citenamefont {Cotton}, \citenamefont
  {Wudarski}, \citenamefont {Urbanek}, \citenamefont {Hait}, \citenamefont
  {Head-Gordon}, \citenamefont {Whaley}, \citenamefont {Moussa}, \citenamefont
  {Wiebe}, \citenamefont {de~Jong},\ and\ \citenamefont
  {Tubman}}]{Klymko2022RealTime}%
  \BibitemOpen
  \bibfield  {author} {\bibinfo {author} {\bibfnamefont {K.}~\bibnamefont
  {Klymko}}, \bibinfo {author} {\bibfnamefont {C.}~\bibnamefont
  {Mejuto-Zaera}}, \bibinfo {author} {\bibfnamefont {S.~J.}\ \bibnamefont
  {Cotton}}, \bibinfo {author} {\bibfnamefont {F.}~\bibnamefont {Wudarski}},
  \bibinfo {author} {\bibfnamefont {M.}~\bibnamefont {Urbanek}}, \bibinfo
  {author} {\bibfnamefont {D.}~\bibnamefont {Hait}}, \bibinfo {author}
  {\bibfnamefont {M.}~\bibnamefont {Head-Gordon}}, \bibinfo {author}
  {\bibfnamefont {K.~B.}\ \bibnamefont {Whaley}}, \bibinfo {author}
  {\bibfnamefont {J.}~\bibnamefont {Moussa}}, \bibinfo {author} {\bibfnamefont
  {N.}~\bibnamefont {Wiebe}}, \bibinfo {author} {\bibfnamefont {W.~A.}\
  \bibnamefont {de~Jong}},\ and\ \bibinfo {author} {\bibfnamefont {N.~M.}\
  \bibnamefont {Tubman}},\ }\href {https://doi.org/10.1103/PRXQuantum.3.020323}
  {\bibfield  {journal} {\bibinfo  {journal} {PRX Quantum}\ }\textbf {\bibinfo
  {volume} {3}},\ \bibinfo {pages} {020323} (\bibinfo {year}
  {2022})}\BibitemShut {NoStop}%
\bibitem [{\citenamefont {Cortes}\ and\ \citenamefont
  {Gray}(2022)}]{Cortes2021QKSD}%
  \BibitemOpen
  \bibfield  {author} {\bibinfo {author} {\bibfnamefont {C.~L.}\ \bibnamefont
  {Cortes}}\ and\ \bibinfo {author} {\bibfnamefont {S.~K.}\ \bibnamefont
  {Gray}},\ }\href {https://doi.org/10.1103/PhysRevA.105.022417} {\bibfield
  {journal} {\bibinfo  {journal} {Physical Review A}\ }\textbf {\bibinfo
  {volume} {105}},\ \bibinfo {pages} {022417} (\bibinfo {year} {2022})},\
  \Eprint {https://arxiv.org/abs/2109.06868} {arXiv:2109.06868 [quant-ph]}
  \BibitemShut {NoStop}%
\bibitem [{\citenamefont {Yoshioka}\ \emph {et~al.}(2025)\citenamefont
  {Yoshioka}, \citenamefont {Amico}, \citenamefont {Kirby}, \citenamefont
  {Jurcevic}, \citenamefont {Dutt}, \citenamefont {Fuller}, \citenamefont
  {Garion}, \citenamefont {Haas}, \citenamefont {Hamamura}, \citenamefont
  {Ivrii}, \citenamefont {Majumdar}, \citenamefont {Minev}, \citenamefont
  {Motta}, \citenamefont {Pokharel}, \citenamefont {Rivero}, \citenamefont
  {Sharma}, \citenamefont {Wood}, \citenamefont {Javadi-Abhari},\ and\
  \citenamefont {Mezzacapo}}]{Yoshioka2025Krylov}%
  \BibitemOpen
  \bibfield  {author} {\bibinfo {author} {\bibfnamefont {N.}~\bibnamefont
  {Yoshioka}}, \bibinfo {author} {\bibfnamefont {M.}~\bibnamefont {Amico}},
  \bibinfo {author} {\bibfnamefont {W.}~\bibnamefont {Kirby}}, \bibinfo
  {author} {\bibfnamefont {P.}~\bibnamefont {Jurcevic}}, \bibinfo {author}
  {\bibfnamefont {A.}~\bibnamefont {Dutt}}, \bibinfo {author} {\bibfnamefont
  {B.}~\bibnamefont {Fuller}}, \bibinfo {author} {\bibfnamefont
  {S.}~\bibnamefont {Garion}}, \bibinfo {author} {\bibfnamefont
  {H.}~\bibnamefont {Haas}}, \bibinfo {author} {\bibfnamefont {I.}~\bibnamefont
  {Hamamura}}, \bibinfo {author} {\bibfnamefont {A.}~\bibnamefont {Ivrii}},
  \bibinfo {author} {\bibfnamefont {R.}~\bibnamefont {Majumdar}}, \bibinfo
  {author} {\bibfnamefont {Z.}~\bibnamefont {Minev}}, \bibinfo {author}
  {\bibfnamefont {M.}~\bibnamefont {Motta}}, \bibinfo {author} {\bibfnamefont
  {B.}~\bibnamefont {Pokharel}}, \bibinfo {author} {\bibfnamefont
  {P.}~\bibnamefont {Rivero}}, \bibinfo {author} {\bibfnamefont
  {K.}~\bibnamefont {Sharma}}, \bibinfo {author} {\bibfnamefont {C.~J.}\
  \bibnamefont {Wood}}, \bibinfo {author} {\bibfnamefont {A.}~\bibnamefont
  {Javadi-Abhari}},\ and\ \bibinfo {author} {\bibfnamefont {A.}~\bibnamefont
  {Mezzacapo}},\ }\href {https://doi.org/10.1038/s41467-025-59716-z} {\bibfield
   {journal} {\bibinfo  {journal} {Nature Communications}\ }\textbf {\bibinfo
  {volume} {16}},\ \bibinfo {pages} {5014} (\bibinfo {year}
  {2025})}\BibitemShut {NoStop}%
\bibitem [{\citenamefont {Yu}\ \emph {et~al.}(2025)\citenamefont {Yu},
  \citenamefont {Robledo~Moreno}, \citenamefont {Iosue}, \citenamefont
  {Bertels}, \citenamefont {Claudino}, \citenamefont {Fuller}, \citenamefont
  {Groszkowski}, \citenamefont {Humble}, \citenamefont {Jurcevic},
  \citenamefont {Kirby}, \citenamefont {Maier}, \citenamefont {Motta},
  \citenamefont {Pokharel}, \citenamefont {Seif}, \citenamefont {Shehata},
  \citenamefont {Sung}, \citenamefont {Tran}, \citenamefont {Tripathi},
  \citenamefont {Mezzacapo},\ and\ \citenamefont {Sharma}}]{Yu2025SKQD}%
  \BibitemOpen
  \bibfield  {author} {\bibinfo {author} {\bibfnamefont {J.}~\bibnamefont
  {Yu}}, \bibinfo {author} {\bibfnamefont {J.}~\bibnamefont {Robledo~Moreno}},
  \bibinfo {author} {\bibfnamefont {J.~T.}\ \bibnamefont {Iosue}}, \bibinfo
  {author} {\bibfnamefont {L.}~\bibnamefont {Bertels}}, \bibinfo {author}
  {\bibfnamefont {D.}~\bibnamefont {Claudino}}, \bibinfo {author}
  {\bibfnamefont {B.}~\bibnamefont {Fuller}}, \bibinfo {author} {\bibfnamefont
  {P.}~\bibnamefont {Groszkowski}}, \bibinfo {author} {\bibfnamefont {T.~S.}\
  \bibnamefont {Humble}}, \bibinfo {author} {\bibfnamefont {P.}~\bibnamefont
  {Jurcevic}}, \bibinfo {author} {\bibfnamefont {W.}~\bibnamefont {Kirby}},
  \bibinfo {author} {\bibfnamefont {T.~A.}\ \bibnamefont {Maier}}, \bibinfo
  {author} {\bibfnamefont {M.}~\bibnamefont {Motta}}, \bibinfo {author}
  {\bibfnamefont {B.}~\bibnamefont {Pokharel}}, \bibinfo {author}
  {\bibfnamefont {A.}~\bibnamefont {Seif}}, \bibinfo {author} {\bibfnamefont
  {A.}~\bibnamefont {Shehata}}, \bibinfo {author} {\bibfnamefont {K.~J.}\
  \bibnamefont {Sung}}, \bibinfo {author} {\bibfnamefont {M.~C.}\ \bibnamefont
  {Tran}}, \bibinfo {author} {\bibfnamefont {V.}~\bibnamefont {Tripathi}},
  \bibinfo {author} {\bibfnamefont {A.}~\bibnamefont {Mezzacapo}},\ and\
  \bibinfo {author} {\bibfnamefont {K.}~\bibnamefont {Sharma}},\ }\href
  {https://doi.org/10.48550/arXiv.2501.09702} {\bibinfo {title}
  {Quantum-centric algorithm for sample-based krylov diagonalization}}
  (\bibinfo {year} {2025}),\ \Eprint {https://arxiv.org/abs/2501.09702}
  {arXiv:2501.09702 [quant-ph]} \BibitemShut {NoStop}%
\bibitem [{\citenamefont {Stair}\ \emph {et~al.}(2020)\citenamefont {Stair},
  \citenamefont {Huang},\ and\ \citenamefont {Evangelista}}]{Stair2020MRSQK}%
  \BibitemOpen
  \bibfield  {author} {\bibinfo {author} {\bibfnamefont {N.~H.}\ \bibnamefont
  {Stair}}, \bibinfo {author} {\bibfnamefont {R.}~\bibnamefont {Huang}},\ and\
  \bibinfo {author} {\bibfnamefont {F.~A.}\ \bibnamefont {Evangelista}},\
  }\href {https://doi.org/10.1021/acs.jctc.9b01125} {\bibfield  {journal}
  {\bibinfo  {journal} {Journal of Chemical Theory and Computation}\ }\textbf
  {\bibinfo {volume} {16}},\ \bibinfo {pages} {2236} (\bibinfo {year}
  {2020})}\BibitemShut {NoStop}%
\bibitem [{\citenamefont {Kirby}\ \emph {et~al.}(2023)\citenamefont {Kirby},
  \citenamefont {Motta},\ and\ \citenamefont {Mezzacapo}}]{Kirby2023Lanczos}%
  \BibitemOpen
  \bibfield  {author} {\bibinfo {author} {\bibfnamefont {W.}~\bibnamefont
  {Kirby}}, \bibinfo {author} {\bibfnamefont {M.}~\bibnamefont {Motta}},\ and\
  \bibinfo {author} {\bibfnamefont {A.}~\bibnamefont {Mezzacapo}},\ }\href
  {https://doi.org/10.22331/q-2023-05-23-1018} {\bibfield  {journal} {\bibinfo
  {journal} {Quantum}\ }\textbf {\bibinfo {volume} {7}},\ \bibinfo {pages}
  {1018} (\bibinfo {year} {2023})}\BibitemShut {NoStop}%
\bibitem [{\citenamefont {Prud'homme}\ \emph {et~al.}(2002)\citenamefont
  {Prud'homme}, \citenamefont {Rovas}, \citenamefont {Veroy}, \citenamefont
  {Machiels}, \citenamefont {Maday}, \citenamefont {Patera},\ and\
  \citenamefont {Turinici}}]{Prudhomme2002ReducedBasis}%
  \BibitemOpen
  \bibfield  {author} {\bibinfo {author} {\bibfnamefont {C.}~\bibnamefont
  {Prud'homme}}, \bibinfo {author} {\bibfnamefont {D.~V.}\ \bibnamefont
  {Rovas}}, \bibinfo {author} {\bibfnamefont {K.}~\bibnamefont {Veroy}},
  \bibinfo {author} {\bibfnamefont {L.}~\bibnamefont {Machiels}}, \bibinfo
  {author} {\bibfnamefont {Y.}~\bibnamefont {Maday}}, \bibinfo {author}
  {\bibfnamefont {A.~T.}\ \bibnamefont {Patera}},\ and\ \bibinfo {author}
  {\bibfnamefont {G.}~\bibnamefont {Turinici}},\ }\href
  {https://doi.org/10.1115/1.1448332} {\bibfield  {journal} {\bibinfo
  {journal} {Journal of Fluids Engineering}\ }\textbf {\bibinfo {volume}
  {124}},\ \bibinfo {pages} {70} (\bibinfo {year} {2002})}\BibitemShut
  {NoStop}%
\bibitem [{\citenamefont {Rammal}\ \emph {et~al.}(2026)\citenamefont {Rammal},
  \citenamefont {Perrin}, \citenamefont {Ladhari}, \citenamefont {Dutreix},
  \citenamefont {Messud},\ and\ \citenamefont
  {Sauban{\`e}re}}]{Rammal2026Orthogonal}%
  \BibitemOpen
  \bibfield  {author} {\bibinfo {author} {\bibfnamefont {H.}~\bibnamefont
  {Rammal}}, \bibinfo {author} {\bibfnamefont {A.}~\bibnamefont {Perrin}},
  \bibinfo {author} {\bibfnamefont {O.}~\bibnamefont {Ladhari}}, \bibinfo
  {author} {\bibfnamefont {C.}~\bibnamefont {Dutreix}}, \bibinfo {author}
  {\bibfnamefont {J.}~\bibnamefont {Messud}},\ and\ \bibinfo {author}
  {\bibfnamefont {M.}~\bibnamefont {Sauban{\`e}re}},\ }\href
  {https://doi.org/10.48550/arXiv.2607.09476} {\bibinfo {title} {Orthogonal
  quantum krylov diagonalisation}} (\bibinfo {year} {2026}),\ \Eprint
  {https://arxiv.org/abs/2607.09476} {arXiv:2607.09476 [quant-ph]} \BibitemShut
  {NoStop}%
\bibitem [{\citenamefont {Asthana}(2025)}]{Asthana2025QKUD}%
  \BibitemOpen
  \bibfield  {author} {\bibinfo {author} {\bibfnamefont {A.}~\bibnamefont
  {Asthana}},\ }\href {https://doi.org/10.48550/arXiv.2512.11788} {\bibinfo
  {title} {Exact and tunable quantum krylov subspaces via unitary
  decomposition}} (\bibinfo {year} {2025}),\ \Eprint
  {https://arxiv.org/abs/2512.11788} {arXiv:2512.11788 [quant-ph]} \BibitemShut
  {NoStop}%
\bibitem [{\citenamefont {Getelina}\ \emph {et~al.}(2024)\citenamefont
  {Getelina}, \citenamefont {Sharma}, \citenamefont {Iadecola}, \citenamefont
  {Orth},\ and\ \citenamefont {Yao}}]{Getelina2024HardwareNoise}%
  \BibitemOpen
  \bibfield  {author} {\bibinfo {author} {\bibfnamefont {J.~C.}\ \bibnamefont
  {Getelina}}, \bibinfo {author} {\bibfnamefont {P.}~\bibnamefont {Sharma}},
  \bibinfo {author} {\bibfnamefont {T.}~\bibnamefont {Iadecola}}, \bibinfo
  {author} {\bibfnamefont {P.~P.}\ \bibnamefont {Orth}},\ and\ \bibinfo
  {author} {\bibfnamefont {Y.-X.}\ \bibnamefont {Yao}},\ }\href
  {https://doi.org/10.1063/5.0217294} {\bibfield  {journal} {\bibinfo
  {journal} {APL Quantum}\ }\textbf {\bibinfo {volume} {1}},\ \bibinfo {pages}
  {036127} (\bibinfo {year} {2024})},\ \Eprint
  {https://arxiv.org/abs/2404.09132} {arXiv:2404.09132 [quant-ph]} \BibitemShut
  {NoStop}%
\bibitem [{\citenamefont {O'Leary}\ \emph {et~al.}(2025)\citenamefont
  {O'Leary}, \citenamefont {Anderson}, \citenamefont {Jaksch},\ and\
  \citenamefont {Kiffner}}]{OLeary2025PQSE}%
  \BibitemOpen
  \bibfield  {author} {\bibinfo {author} {\bibfnamefont {T.}~\bibnamefont
  {O'Leary}}, \bibinfo {author} {\bibfnamefont {L.~W.}\ \bibnamefont
  {Anderson}}, \bibinfo {author} {\bibfnamefont {D.}~\bibnamefont {Jaksch}},\
  and\ \bibinfo {author} {\bibfnamefont {M.}~\bibnamefont {Kiffner}},\ }\href
  {https://doi.org/10.22331/q-2025-05-05-1726} {\bibfield  {journal} {\bibinfo
  {journal} {Quantum}\ }\textbf {\bibinfo {volume} {9}},\ \bibinfo {pages}
  {1726} (\bibinfo {year} {2025})}\BibitemShut {NoStop}%
\bibitem [{\citenamefont {De~Vriendt}\ \emph {et~al.}(2026)\citenamefont
  {De~Vriendt}, \citenamefont {Bringewatt}, \citenamefont {Gjonbalaj},
  \citenamefont {Ostermann}, \citenamefont {Vodola}, \citenamefont
  {Borregaard}, \citenamefont {K{\"u}hn},\ and\ \citenamefont
  {Yelin}}]{DeVriendt2026Angle}%
  \BibitemOpen
  \bibfield  {author} {\bibinfo {author} {\bibfnamefont {X.}~\bibnamefont
  {De~Vriendt}}, \bibinfo {author} {\bibfnamefont {J.}~\bibnamefont
  {Bringewatt}}, \bibinfo {author} {\bibfnamefont {N.~O.}\ \bibnamefont
  {Gjonbalaj}}, \bibinfo {author} {\bibfnamefont {S.}~\bibnamefont
  {Ostermann}}, \bibinfo {author} {\bibfnamefont {D.}~\bibnamefont {Vodola}},
  \bibinfo {author} {\bibfnamefont {J.}~\bibnamefont {Borregaard}}, \bibinfo
  {author} {\bibfnamefont {M.}~\bibnamefont {K{\"u}hn}},\ and\ \bibinfo
  {author} {\bibfnamefont {S.~F.}\ \bibnamefont {Yelin}},\ }\href
  {https://doi.org/10.48550/arXiv.2602.11985} {\bibinfo {title} {A new angle on
  quantum subspace diagonalization for quantum chemistry}} (\bibinfo {year}
  {2026}),\ \Eprint {https://arxiv.org/abs/2602.11985} {arXiv:2602.11985
  [quant-ph]} \BibitemShut {NoStop}%
\bibitem [{\citenamefont {Lee}\ \emph {et~al.}(2024)\citenamefont {Lee},
  \citenamefont {Lee},\ and\ \citenamefont {Huh}}]{Lee2024Sampling}%
  \BibitemOpen
  \bibfield  {author} {\bibinfo {author} {\bibfnamefont {G.}~\bibnamefont
  {Lee}}, \bibinfo {author} {\bibfnamefont {D.}~\bibnamefont {Lee}},\ and\
  \bibinfo {author} {\bibfnamefont {J.}~\bibnamefont {Huh}},\ }\href
  {https://doi.org/10.22331/q-2024-09-19-1477} {\bibfield  {journal} {\bibinfo
  {journal} {Quantum}\ }\textbf {\bibinfo {volume} {8}},\ \bibinfo {pages}
  {1477} (\bibinfo {year} {2024})}\BibitemShut {NoStop}%
\bibitem [{\citenamefont {Kirby}(2024)}]{Kirby2024Errors}%
  \BibitemOpen
  \bibfield  {author} {\bibinfo {author} {\bibfnamefont {W.}~\bibnamefont
  {Kirby}},\ }\href {https://doi.org/10.22331/q-2024-08-29-1457} {\bibfield
  {journal} {\bibinfo  {journal} {Quantum}\ }\textbf {\bibinfo {volume} {8}},\
  \bibinfo {pages} {1457} (\bibinfo {year} {2024})}\BibitemShut {NoStop}%
\bibitem [{\citenamefont {Zhang}\ \emph {et~al.}(2024)\citenamefont {Zhang},
  \citenamefont {Wang}, \citenamefont {Xu},\ and\ \citenamefont
  {Li}}]{Zhang2024MeasEff}%
  \BibitemOpen
  \bibfield  {author} {\bibinfo {author} {\bibfnamefont {Z.}~\bibnamefont
  {Zhang}}, \bibinfo {author} {\bibfnamefont {A.}~\bibnamefont {Wang}},
  \bibinfo {author} {\bibfnamefont {X.}~\bibnamefont {Xu}},\ and\ \bibinfo
  {author} {\bibfnamefont {Y.}~\bibnamefont {Li}},\ }\href
  {https://doi.org/10.22331/q-2024-08-13-1438} {\bibfield  {journal} {\bibinfo
  {journal} {Quantum}\ }\textbf {\bibinfo {volume} {8}},\ \bibinfo {pages}
  {1438} (\bibinfo {year} {2024})}\BibitemShut {NoStop}%
\bibitem [{\citenamefont {Lee}\ \emph {et~al.}(2025)\citenamefont {Lee},
  \citenamefont {Choi}, \citenamefont {Huh},\ and\ \citenamefont
  {Izmaylov}}]{Lee2025Sampling}%
  \BibitemOpen
  \bibfield  {author} {\bibinfo {author} {\bibfnamefont {G.}~\bibnamefont
  {Lee}}, \bibinfo {author} {\bibfnamefont {S.}~\bibnamefont {Choi}}, \bibinfo
  {author} {\bibfnamefont {J.}~\bibnamefont {Huh}},\ and\ \bibinfo {author}
  {\bibfnamefont {A.~F.}\ \bibnamefont {Izmaylov}},\ }\href
  {https://doi.org/10.1039/D4DD00321G} {\bibfield  {journal} {\bibinfo
  {journal} {Digital Discovery}\ }\textbf {\bibinfo {volume} {4}},\ \bibinfo
  {pages} {954} (\bibinfo {year} {2025})}\BibitemShut {NoStop}%
\bibitem [{\citenamefont {Oliveira}\ \emph {et~al.}(2026)\citenamefont
  {Oliveira}, \citenamefont {Ziems},\ and\ \citenamefont
  {Glaser}}]{Oliveira2026Instabilities}%
  \BibitemOpen
  \bibfield  {author} {\bibinfo {author} {\bibfnamefont {M.~G.~J.}\
  \bibnamefont {Oliveira}}, \bibinfo {author} {\bibfnamefont {K.~M.}\
  \bibnamefont {Ziems}},\ and\ \bibinfo {author} {\bibfnamefont
  {N.}~\bibnamefont {Glaser}},\ }\href
  {https://doi.org/10.48550/arXiv.2604.11532} {\bibinfo {title} {Tackling
  instabilities of quantum krylov subspace methods: An analysis of the
  numerical and statistical errors}} (\bibinfo {year} {2026}),\ \Eprint
  {https://arxiv.org/abs/2604.11532} {arXiv:2604.11532 [quant-ph]} \BibitemShut
  {NoStop}%
\bibitem [{\citenamefont {Xu}\ \emph {et~al.}(2026)\citenamefont {Xu},
  \citenamefont {Shirakawa},\ and\ \citenamefont
  {Yunoki}}]{Xu2026FilterAssisted}%
  \BibitemOpen
  \bibfield  {author} {\bibinfo {author} {\bibfnamefont {H.}~\bibnamefont
  {Xu}}, \bibinfo {author} {\bibfnamefont {T.}~\bibnamefont {Shirakawa}},\ and\
  \bibinfo {author} {\bibfnamefont {S.}~\bibnamefont {Yunoki}},\ }\href
  {https://doi.org/10.48550/arXiv.2605.28040} {\bibinfo {title}
  {Filter-assisted quantum subspace diagonalization via wavefunction sparsity
  engineering}} (\bibinfo {year} {2026}),\ \Eprint
  {https://arxiv.org/abs/2605.28040} {arXiv:2605.28040 [quant-ph]} \BibitemShut
  {NoStop}%
\bibitem [{\citenamefont {Verstraete}\ \emph {et~al.}(2009)\citenamefont
  {Verstraete}, \citenamefont {Wolf},\ and\ \citenamefont
  {Cirac}}]{Verstraete2009Dissipation}%
  \BibitemOpen
  \bibfield  {author} {\bibinfo {author} {\bibfnamefont {F.}~\bibnamefont
  {Verstraete}}, \bibinfo {author} {\bibfnamefont {M.~M.}\ \bibnamefont
  {Wolf}},\ and\ \bibinfo {author} {\bibfnamefont {J.~I.}\ \bibnamefont
  {Cirac}},\ }\href {https://doi.org/10.1038/nphys1342} {\bibfield  {journal}
  {\bibinfo  {journal} {Nature Physics}\ }\textbf {\bibinfo {volume} {5}},\
  \bibinfo {pages} {633} (\bibinfo {year} {2009})}\BibitemShut {NoStop}%
\bibitem [{\citenamefont {Li}\ \emph {et~al.}(2025)\citenamefont {Li},
  \citenamefont {Zhan},\ and\ \citenamefont {Lin}}]{Li2025Dissipative}%
  \BibitemOpen
  \bibfield  {author} {\bibinfo {author} {\bibfnamefont {H.-E.}\ \bibnamefont
  {Li}}, \bibinfo {author} {\bibfnamefont {Y.}~\bibnamefont {Zhan}},\ and\
  \bibinfo {author} {\bibfnamefont {L.}~\bibnamefont {Lin}},\ }\href
  {https://doi.org/10.1038/s41534-025-01124-8} {\bibfield  {journal} {\bibinfo
  {journal} {npj Quantum Information}\ }\textbf {\bibinfo {volume} {11}},\
  \bibinfo {pages} {183} (\bibinfo {year} {2025})}\BibitemShut {NoStop}%
\bibitem [{\citenamefont
  {Herrera-Mart{\'i}}(2026)}]{HerreraMarti2026OpenKrylov}%
  \BibitemOpen
  \bibfield  {author} {\bibinfo {author} {\bibfnamefont {D.~A.}\ \bibnamefont
  {Herrera-Mart{\'i}}},\ }\href {https://doi.org/10.48550/arXiv.2605.03715}
  {\bibinfo {title} {Real-time krylov diagonalisation for open quantum
  systems}} (\bibinfo {year} {2026}),\ \Eprint
  {https://arxiv.org/abs/2605.03715} {arXiv:2605.03715 [quant-ph]} \BibitemShut
  {NoStop}%
\bibitem [{\citenamefont {Dalibard}\ \emph {et~al.}(1992)\citenamefont
  {Dalibard}, \citenamefont {Castin},\ and\ \citenamefont
  {M{\o}lmer}}]{Dalibard1992WaveFunction}%
  \BibitemOpen
  \bibfield  {author} {\bibinfo {author} {\bibfnamefont {J.}~\bibnamefont
  {Dalibard}}, \bibinfo {author} {\bibfnamefont {Y.}~\bibnamefont {Castin}},\
  and\ \bibinfo {author} {\bibfnamefont {K.}~\bibnamefont {M{\o}lmer}},\ }\href
  {https://doi.org/10.1103/PhysRevLett.68.580} {\bibfield  {journal} {\bibinfo
  {journal} {Physical Review Letters}\ }\textbf {\bibinfo {volume} {68}},\
  \bibinfo {pages} {580} (\bibinfo {year} {1992})}\BibitemShut {NoStop}%
\end{thebibliography}%

\end{document}